\documentclass[12pt]{article}
\pdfoutput=1
\usepackage{jcapmod}
\usepackage{booktabs}
\usepackage{mathtools}
\usepackage[english]{babel}
\usepackage{amsmath,amssymb,amsbsy,amstext, amsthm}
\usepackage{hyperref}
\usepackage[usenames,dvipsnames]{xcolor}
\hypersetup{
    urlcolor=NavyBlue,
    citecolor=NavyBlue,
    linkcolor=NavyBlue,
}%
\usepackage{array}
\usepackage{url} 
\usepackage{slashed}
\usepackage{multicol}
\usepackage{blkarray}
\usepackage{enumerate}
\usepackage{graphicx}
\usepackage{amsfonts}
\usepackage{amssymb}
  \usepackage{empheq}
  \usepackage{tcolorbox}
\usepackage{cleveref}
\usepackage{colortbl}
  \newcommand*\widefbox[1]{\fbox{\hspace{2em}#1\hspace{2em}}}
\usepackage{float}
\usepackage[utf8]{inputenc}
\usepackage{soul}
\usepackage[normalem]{ulem}
\RequirePackage{color}
\usepackage{colortbl}
\definecolor{green2}{cmyk}{0, 1, 0.5, 0}
\definecolor{lightgreen}{cmyk}{0.2, 0, 0.2, 0.2}
\definecolor{lightgray}{cmyk}{0.1,0.2,0,0.1}
\definecolor{lightgray2}{cmyk}{0.4,0.4,0,0.8}
\definecolor{black}{cmyk}{1.0,1.0,1.0,1.0}

\allowdisplaybreaks[1]

\usepackage{colortbl}
\definecolor{lightgreen}{cmyk}{0.2, 0, 0.2, 0.2}
\definecolor{lightgray}{cmyk}{0.1,0.2,0,0.1}
\definecolor{lightgray2}{cmyk}{0.1,0.1,0,0.1}

\makeatletter
\newlength{\apb@width}
\newcommand{\autoparbox}[2][c]{\settowidth{\apb@width}{#2}\parbox[#1]{\apb@width}{#2}}

\makeatother

\numberwithin{equation}{section}

\def\be{\begin{equation}}
\def\ee{\end{equation}}

\def\bea{\begin{eqnarray}}
\def\eea{\end{eqnarray}}

\def\lp{\left(}
\def\rp{\right)}

\def\lb{\left[}
\def\rb{\right]}

\def\d{{\rm d}}

\def\d{{\rm d}}

\def\del{\partial}

\def\0{{\boldsymbol 0}}

\def\omk{\Omega_k}

\usepackage{setspace} 
\begin{document}
%\linespread{1.5}

%\doublespacing
\begin{titlepage}

\setcounter{page}{1} \baselineskip=15.5pt \thispagestyle{empty}

\bigskip\

\vspace{1cm}
\begin{center}

{\fontsize{20}{28}\selectfont  \sffamily \bfseries {Cosmological constraints on curved quintessence
 \\}}

\end{center}

\vspace{0.2cm}

\begin{center}
{\fontsize{13}{30}\selectfont Sukannya Bhattacharya$^{a,}$\footnote{\texttt{sukannya.bh@gmail.com}}, Giulia Borghetto$^{b,\!}$
\footnote{\texttt{giulia.borghetto@gmail.com}}, Ameek Malhotra$^{b,\!}$ \footnote{\texttt{ameek.malhotra@swansea.ac.uk}}, \\
Susha Parameswaran$^{c,}$\footnote{\texttt{susha.parameswaran@liverpool.ac.uk}},
Gianmassimo Tasinato$^{b,d,}$\footnote{\texttt{g.tasinato2208@gmail.com}},  Ivonne Zavala$^{b,}$\footnote{\texttt{e.i.zavalacarrasco@swansea.ac.uk}}
} 
\end{center}

\begin{center}

\vskip 8pt
\textsl{$^a$ Dipartimento di Fisica e Astronomia ``G. Galilei'', Universit\`a degli Studi di Padova, \\
Via Marzolo 8, 35131 Padova, Italy}\\
\textsl{$^b$ Physics Department, Swansea University, SA2 8PP, UK}\\
\textsl{$^c$ Department of Mathematical Sciences, University of Liverpool, \\ Liverpool, 
L69 7ZL, United Kingdom}\\
\textsl{$^d$ Dipartimento di Fisica e Astronomia, Universit\`a di Bologna and \\
 INFN, Sezione di Bologna, I.S. FLAG, viale B. Pichat 6/2, 40127 Bologna,   Italy
}
\vskip 6pt

\end{center}

\vspace{1.2cm}
\hrule \vspace{0.3cm}
\noindent
\noindent
Dynamical dark energy has  gained renewed interest due to recent theoretical and observational developments.  In the present paper, we  focus on a string-motivated dark energy set-up, and perform a detailed cosmological analysis of exponential quintessence  with potential  $V=V_0 e^{-\lambda\phi}$, allowing for non-zero spatial curvature. 
We first gain some physical intuition into the full evolution of such a scenario by analysing the corresponding dynamical system.  Then, we  test the model using  a combination of  \textit{Planck} CMB data, DESI BAO data, as well as recent supernovae datasets. For the model parameter $\lambda$, we obtain a  preference for nonzero values: \mbox{$\lambda = 0.48^{+0.28}_{-0.21},\; 0.68^{+0.31}_{-0.20},\; 0.77^{+0.18}_{-0.15}$} at 68\% C.L.~when combining CMB+DESI with Pantheon+, Union3 and DES-Y5 supernovae datasets respectively.
We find no significant hint for spatial curvature. 
We discuss the implications of current cosmological results for the exponential quintessence model, and  more generally for dark  energy in string theory. 
\vskip 10pt
\hrule

\vspace{0.4cm}
 \end{titlepage}

 \tableofcontents

%\newpage

\section{Introduction}

The $\Lambda$CDM model of cosmology has emerged in the last 20+ years as a successful  phenomenological model to describe the evolution of our universe from the time of the primordial Big-Bang to today.  In this scenario, the present accelerated expansion, discovered at the turn of the 21st century \cite{SupernovaSearchTeam:1998fmf,SupernovaCosmologyProject:1998vns}, is described by a positive cosmological constant, $\Lambda$, while the three-dimensional space curvature is taken to be flat.   In  recent years, thanks to the explosive increase in the volume and accuracy of cosmological measurements, challenges to this model have started to emerge (see \cite{Abdalla:2022yfr} for a recent summary of the so-called cosmological tensions).
% the most notable being the $H_0$-tension between early and late time measurements. 
This is rather good news, as it places us in a remarkable era where fundamental models can be tested and a more accurate cosmological model can emerge.  
 In this work, we
study string-motivated dark energy models and confront them with
%at the light of 
 the most recent cosmological data. %, including  DESI. 

\smallskip

Indeed, from the theoretical point of view, much work has been focused on a better understanding of the present (and early) acceleration of the universe and the nature of its source, dark energy, modelled by a cosmological constant in the $\Lambda$CDM.  Since soon after its discovery, cosmic  acceleration has caused concerns due to the  appearance  of cosmological horizons in eternal acceleration and questions around the consistency of such horizons in a theory of quantum gravity, specifically in string theory  \cite{Fischler:2001yj,Hellerman:2001yi}.  The possibility of realising a de Sitter (dS) universe, namely a universe dominated by a positive cosmological constant, within string theory has subsequently been a source of extensive work and debate (see e.g.~\cite{Cicoli:2023opf} for a review). In particular, it has recently been provocatively suggested that dS space cannot arise in string theory and more generally, in quantum gravity \cite{Garg:2018reu,Ooguri:2018wrx}.  

The possibility of a dynamical form of energy, rather than a constant, was put forward early on after the discovery of the late time acceleration, and has received continuous attention thereafter.  The main such candidate is {\em quintessence}, a scalar field whose potential energy drives the accelerated expansion  \cite{Ratra:1987rm,Peebles:1987ek,Caldwell:1997ii}.
In particular, quintessence models governed by exponential potentials \cite{Copeland:1997et,Cline:2001nq,Kolda:2001ex,Agrawal:2018own}, $V=V_0 \,e^{-\lambda\phi}$, are especially interesting as they are ubiquitous at the boundaries of moduli spaces in string theory, which are regions that are parametrically under good theoretical control. 
Cosmological constraints on  exponential quintessence with flat spatial slices have been studied recently  in~\cite{Akrami:2018ylq,Raveri:2018ddi,Schoneberg:2023lun}, where upper bounds on the exponent $\lambda$ were obtained using a combination of \textit{Planck} CMB data, BAO data and supernovae data available at that time. 
Moreover, in the recent study~\cite{Schoneberg:2023lun},  cosmological constraints of other quintessence potentials were also analysed under the assumption of flat spatial geometry. 
 
Meanwhile, it has been highlighted recently \cite{Andriot:2023wvg} (see also \cite{Boya:2002mv}) that including spatial curvature in the context of exponential quintessence may allow for eternally accelerating solutions {\it{without cosmological horizons}} for exponents that are well motivated by string theory constructions, $\lambda>\sqrt{2}$.  It has also been argued that an open universe is a natural outcome of  Coleman-de Luccia tunnelling in the string landscape \cite{Freivogel:2005vv}, although other possibilities may also exist, as recently investigated in e.g.~\cite{Cespedes:2020xpn,Cespedes:2023jdk}.
As a further motivation to consider spatial curvature,  a recent study \cite{Bousso:2022gth} (see also \cite{Ben-Dayan:2022nmb})
in the context of  {\em islands in cosmology}\footnote{Islands have been proposed  in the context of the black hole information paradox as hypothetical regions inside the black hole that help encode information from it, ensuring it is not lost
in evaporation (see \cite{Almheiri:2020cfm} for a  review). The boundary of such island is called quantum extremal surface (QES) \cite{Engelhardt:2014gca}.}  \cite{Hartman:2020khs} to understand holography beyond adS/CFT, has found that a small amount of spatial curvature can
have a significant effect in the appearance of islands. In particular, in the presence of negative, positive or zero cosmological constant, arbitrarily small positive curvature allows  the entire spacetime to be an island. On the other hand, a small amount of negative curvature eliminates cosmological islands entirely.

The full dynamical system including radiation, matter, quintessence and negative spatial curvature was studied in detail not long ago 
in \cite{Andriot:2024jsh} (see \cite{Bahamonde:2017ize} for a review of dynamical systems in cosmology and e.g. \cite{vandenHoogen:1999qq, Gosenca:2015qha, Marconnet:2022fmx, Andriot:2023wvg, SavasArapoglu:2017pyh} for 
some 
studies of relevant subsystems).   An upper bound, $\lambda \lesssim \sqrt{3}$, was found from the minimal phenomenological requirements of a past epoch of radiation domination and acceleration today.  Cosmological solutions were shown to start universally in the past from the unique fully unstable fixed-point, a kination epoch, then pass through radiation and matter dominated phases and an epoch of acceleration as finally the unique stable fixed-point is approached.  
The characteristics of the attractor fixed-point depend on $\lambda$; for $\lambda>\sqrt{2}$ it corresponds to a curvature scaling solution, with equation of state $w=-\frac13$ whilst for $\lambda \leq \sqrt{2}$ it corresponds to scalar domination with $w<-1/3$.  How the attractor is approached depends on the initial conditions and consequent trajectory; it turns out that for $\lambda>\sqrt{2}$ and past radiation and matter domination the epoch of acceleration is only transient, whereas for $\lambda \leq \sqrt{2}$ the acceleration is eternal and there is an  associated cosmological horizon. 

Alongside these theoretical developments and motivations, from the cosmological point of view, determining the equation of state of dark energy has been one of the main research motivations and aims of  several observational missions such as DES \cite{DES:2024tys}, DESI \cite{DESI:2024lzq}, Euclid \cite{Euclid:2019clj} and LSST \cite{2009arXiv0912.0201L}. 
Indeed, recent data from the DES \cite{DES:2024tys} and DESI \cite{DESI:2024lzq,DESI:2024uvr,DESI:2024mwx,Calderon:2024uwn,DESI:2024kob,Lodha:2024upq}
surveys hint at a preference for a dynamical dark energy.\footnote{See~\cite{Colgain:2024xqj,Carloni:2024zpl,Park:2024jns,Wang:2024rjd,Cortes:2024lgw,Wang:2024pui,Dinda:2024kjf,Croker:2024jfg,Wang:2024hks,Luongo:2024fww,Mukherjee:2024ryz,Wang:2024dka} for recent discussions related to the interpretations of the DESI results. Refs~\cite{Tada:2024znt,Yin:2024hba,Berghaus:2024kra,Shlivko:2024llw} have also analysed the viability of quintessence models in light of the DESI findings.} 

These surveys use the so called $w_0w_a$ or Chevallier-Polarski-Linder (CPL) parameterisation \cite{Chevallier:2000qy,Linder:2002et}  of the equation of state $w$ (at least for  small redshifts), which varies linearly with the scale factor $a$, that is
\be\label{eq:w0wa}
w(a) = w_0+ w_a(1-a)\,.
\ee
The exponential quintessence model in the presence of spatial curvature has not yet been tested against the most recent cosmological data (see \cite{Aurich:2003it} for an early study using WMAP), in particular the late time probes such as BAO from DESI~\cite{DESI:2024mwx}, and supernovae datasets such as DES~\cite{DES:2024tys}, Pantheon+~\cite{Brout:2022vxf} and Union3~\cite{Rubin:2023ovl}. 

Therefore, we perform a cosmological analysis of  exponential quintessence models in curved space   ($k=\pm1$) using the latest cosmological data (see also Appendix \ref{app:full_constraints} for the flat case $k=0$).  
In what follows, we refer to this model as ``curved quintessence". When we discuss this model in the context of cosmological constraints, we refer to it as $q$CDM+$\Omega_k$ (or simply $q$CDM). As we will see, including the most recent data from DESI and supernovae data from DES, Pantheon+ and Union3, we are able to constrain the parameter $\lambda$, finding it to lie roughly $2\text{--}4\sigma$ away from $\lambda=0$ (which represents the cosmological constant), depending on the supernovae dataset chosen. We quantify the improvement in the fit to data provided by the $q$CDM model as compared to $\Lambda$CDM and also compare this model against the CPL parametrisation, finding a mild preference for the latter. 

%
%We organise the paper as follows. 
In section \ref{sec:cosmo} we introduce the cosmological system and perform a dynamical system analysis in \ref{sec:dynsys} for the closed case $k=+1$, while the open case $k=-1$ is reproduced from \cite{Andriot:2024jsh} in Appendix \ref{app1} (the flat case can be found in \cite{Bahamonde:2017ize}). In section \ref{sec:constrains} we perform the full cosmological analysis of the model with exponential potential for quintessence with curved spacetime geometry. We preset the constraints on the relevant model parameters and discuss their implications.
The constraints on the full set of parameters, including 6 base parameters in the standard $\Lambda$CDM model, and 2 more for the quintessence model and curvature, are presented in Appendix \ref{app:full_constraints}.
Finally in section~\ref{sec:discussion} we summarise our findings. 

See \cite{Monteroetal} for an independent cosmological analysis of curved quintessence and~\cite{Ramadan:2024kmn} for the analysis in the flat case.

\section{Cosmological system}
\label{sec:cosmo}

We start 
 by introducing the cosmological system we are interested in analysing:  quintessence \cite{Ratra:1987rm,Peebles:1987ek,Caldwell:1997ii} as a possible description of dark energy (DE), together with matter and radiation in a curved space.  This constitutes a modification of the standard $\Lambda$CDM model, where the cosmological constant $\Lambda$, is replaced by a (canonically normalised) scalar field $\phi$, and we allow for curvature ($k\neq 0$) of the 3D space slices. 
 The corresponding
dynamical analysis is rich in possibilities, which
we analyze in this section. In the next sections we study how current data are able to discriminate among different options. 

We consider a 4D Friedmann Lema\^itre Robertson Walker (FLRW) metric  with arbitrary curvature given by:
\be\label{metric}
ds^2 = -dt^2 +a^2(t) \left[\frac{dr^2}{1-kr^2}+r^2\left(d\theta ^2 +\sin ^2\theta d\psi ^2\right)\right] \,,
\ee
where $k=0,\pm1$ denotes the curvature of the three-dimensional (3D) slices. 
The radiation ($r$), matter ($m$) and quintessence field are each described by perfect fluids with energy density, $\rho_i$, and pressure, $p_i$, related by their {\em equation of state} parameter, $w_i$ as:
\be
p_i=w_i\rho_i\,,
\ee
with $i=r, m, \phi$, respectively. For radiation, $\rho_r \sim a^{-4}$ and $w_r=\frac13$; for matter, $\rho_m \sim a^{-3}$ and $w_m=0$; for the scalar, the energy density and pressure are given by 
\be\label{eq:rhofi}
\rho_\phi = \frac{\dot\phi^2}{2} + V(\phi) \,,\qquad p_\phi = \frac{\dot\phi^2}{2} - V(\phi)\,.
\ee
Moreover, we can introduce an effective ``curvature fluid component'', with energy density, pressure and equation of state given by
\be\label{eq:rhok}
\rho_k = -\frac{3\,k}{a^2}\,,  \qquad  p_k = \frac{k}{a^2} \,, \qquad w_k = -\frac{1}{3} \,.
\ee

The equations of motion for this system are given by (we set $M_{\rm Pl}^{-2}=8\pi G=1$ in this section):
\begin{subequations}\label{eq:eoms}
 \begin{align}
      H^2 &= \frac{\rho_{\rm eff}}{3}\,,
      %\left(\rho_r +\rho_m+ \rho_\phi +\rho_k\right) 
      \label{eq:Fried}\\
   - \frac{ \dot H}{H^2} &
    %= -\frac{1}{2H^2}
    %\left[\rho_\phi(1+w_\phi) +\rho(1+\omega) + \rho_k(1+\omega_k)\right] 
    %\sum_n\rho_n(1+w_n)
    =\frac{3}{2}(1+w_{\rm eff})\,,
    %\frac{1}{2H^2}\rho_{\rm eff}(1+w_{\rm eff})
    \label{eq:Hdot}\\
    \ddot \phi &= -3H\dot \phi -V_\phi \,.  \label{eq:phi}
 \end{align}
\end{subequations}
In these equations, $V_\phi\equiv \del_{\phi} V$ and we  defined
\be
\rho_{\rm eff} = \sum_n \rho_n\,, \qquad p_{\rm eff} = \sum_n p_n\,, \qquad p_{\rm eff}=w_{\rm eff}\, \rho_{\rm eff}.
\ee
with $n=r,m, \phi, k$, that is, running over radiation, matter, the scalar  and curvature terms. From this definition we can deduce  that 
\be
w_{\rm eff} = \sum_n w_n\Omega_n\,.
\ee
Moreover, from \eqref{eq:Hdot}, one can check that
cosmic acceleration requires
\be\label{eq:eps}
w_{\rm eff}<-1/3 \quad \Leftrightarrow \quad \epsilon\equiv -\dot H/H^2<1\,. 
\ee
Introducing the density parameters for each component, $n$, in the universe as 
\be
\Omega_n=\frac{\rho_n}{3H^2}\,,
\ee
we can write the first Friedmann equation as 
\bea
\label{eq:friO}
1 = \sum_n\Omega_n\,, \qquad %\nonumber\\
\text{or} \qquad 1-\Omega_k =\Omega_T,
\eea
where 
\be
\Omega_T = \sum_i\Omega_i\,.
%\,,\qquad \Omega_k = - \frac{k}{a^2 H^2} \,.
\ee
From \eqref{eq:friO}, we see the standard result that in an open universe ($k=-1$), $\Omega_T<1$,  in a closed universe ($k=1$), $\Omega_T>1$ and in a flat universe $\Omega_T=1$. 

For the rest of our
analysis we restrict ourselves to an exponential  potential for the scalar field, given by 
\be
V(\phi) = V_0\ e^{-\lambda \phi} \ , \label{potexp}
\ee
where $\lambda$ is a constant and we take $\lambda>0$, $V_0\geq0$.  As discussed
in the Introduction, this is motivated from the general form of a perturbatively generated potential for a closed string modulus after canonical normalisation (see e.g. \cite{Cicoli:2023opf}).

\subsection{Dynamical systems analysis}\label{sec:dynsys}
To study the background evolution  of the system, it is useful to translate the equations of motion into an autonomous system of first order coupled differential equations, and perform a  dynamical systems analysis. Curved exponential quintessence  including a barotropic fluid has been studied previously in \cite{vandenHoogen:1999qq} and in \cite{Gosenca:2015qha} an analysis was performed including matter and radiation (see also \cite{Bahamonde:2017ize} for a  review on dynamical systems in cosmology and references therein). 
More recently, in \cite{Andriot:2024jsh} the open case ($k=-1$) including radiation and matter was studied in  detail. 
 We refer the reader to that paper for details and present a summary of their main results in  Appendix \ref{app1}.
 Below we give some details on the closed case ($k=+1$). 
As pointed out in \cite{vandenHoogen:1999qq,Bahamonde:2017ize}, the dynamical system in the closed case is non-compact. 

We  define the following  dynamical system variables:
\be\label{eq:variablesu}
 \bar x= \frac{\phi'}{\sqrt{6}} \,, \qquad 
   \bar y = \frac{\sqrt{V}}{\sqrt{3} \bar H}\,,  
   \qquad \bar u =\frac{\sqrt{\rho_r}}{\sqrt{3}\bar H}\,,
   \qquad  \bar z=\frac{H}{\bar H} \,, \qquad   
   \lambda = -\frac{V_\phi}{V} \,, 
   \ee
   together with the constraint 
   \be\label{eq:Omegamxyu}
   \bar\Omega_m=1-\bar x^2-\bar y^2-\bar u^2\,,
   \ee
where 
\be\label{eq:barOm}
  \bar\Omega_m \equiv \frac{\rho_m}{3\overline{H}^2}\,,
\ee   
and, importantly, we  introduce 
\be
\bar{H} =H \sqrt{1-\Omega_k}\,, \quad \text{where} \quad\Omega_k\equiv-\frac{k}{a^2H^2} \,, \quad \text{and} \quad k> 0\,, 
\ee
%\frac{k}{a^2H^2}}
to compactify our phase space \cite{Bahamonde:2017ize}.
Primes above denote derivatives with respect to the new e-fold variable:
\be
d\overline{N} =  \overline{H} dt \,.
\ee
The equations of motion \eqref{eq:eoms} can  be rewritten in terms of  variables \eqref{eq:variablesu} as follows:
\begin{subequations}\label{eq:system2}
 \begin{empheq}[box=\widefbox]{align}
    \bar{x}'&= \frac12\,\bar x\,\bar z \lp -3+\bar u^2 +3\bar x^2 -3\bar y^2\rp +  \sqrt{\frac{3}{2}} \lambda \,\bar y^2  \,,
    \label{eq:xp2}\\
  \bar{y}' &= \frac12\, \bar y\,\bar z\lp 3+\bar u^2 +3\bar x^2 -3\bar y^2 \rp - \sqrt{\frac{3}{2}}\,\lambda  \, \bar x \,\bar y\, \,, \label{eq:yp2}
  \\
  \bar{u}' &= \frac12  \,\bar u\,\bar z \lp -1+\bar u^2+3\bar x^2-3\bar y^2 \rp  \,, \label{eq:up2}
  \\
  \bar{z}' & =\frac{1}{2}\lp \bar{z}^2 -1\rp\lp 1+\bar u^2+3\bar{x}^2 -3 \bar{y}^2 \rp \,, \label{eq:zp2} \\
        \lambda' &= -\sqrt{6}\,\bar{x}\, \lb \frac{V_{\phi\phi}}{V} -\frac{V_{\phi}^2}{V^2}\rb\label{eq:lamb2}\,.
\end{empheq}
\end{subequations}
%
%where  a prime now indicates derivatives with respect to $\overline{N}$. 

The fixed points for this system are given in  table  \ref{tab:2} below; since we are interested only in expanding universes, we stick there to the case $\bar z> 0$. The stability of the closed universe fixed points is summarised in table \ref{tab:3}.
For comparison, the fixed points for the open case as well as their properties and stability as discussed in  \cite{Andriot:2024jsh} are reproduced in %table \ref{tab:1} in 
Appendix \ref{app1}.

\begin{table}[h]
\begin{center}
\centering
\begin{tabular}{| l | c | c | c | }
\hline
\cellcolor[gray]{0.9} & \cellcolor[gray]{0.9} & \cellcolor[gray]{0.9} & \cellcolor[gray]{0.9}\\[-8pt]
\cellcolor[gray]{0.9} \hskip 1.4cm $(\bar x, \bar y, \bar z, \bar u)$ & \cellcolor[gray]{0.9} $\bar\Omega_m$ &  \cellcolor[gray]{0.9} Existence & \cellcolor[gray]{0.9} $w_{\rm eff}$ \\[5pt]
\hline
&&&\\[-8pt]
$\bar Q_{\rm kin}^\pm= $ $(\pm 1,0,1,0) $ & $0$  & $\forall\, \lambda$ & $1$ \\[3pt]
\hline
&&&\\[-8pt]
$\bar Q_{m}= $ $(0,0,1,0) $ & $1$  & $ \forall\, \lambda$ & $ 0$ \\[3pt]
\hline
&&&\\[-8pt]
$\bar Q_{k\,\phi}= $ $\lp \frac{1}{\sqrt{3}},\pm\sqrt{\frac{2}{3}} ,\frac{\lambda}{\sqrt{2}},0\rp $ & $0$ & $ \lambda<\sqrt{2}$ & $ -\frac{1}{3}$ \\[8pt]
%&&&\\[-8pt]
 &  & (For $  \lambda = \sqrt{2}$, $\bar Q_{k\phi}= \bar Q_{\phi}$)  & \\[8pt]
\hline
&&&\\[-8pt]
$\bar Q_{\phi}= $ $\lp \frac{\lambda}{\sqrt{6}}, \pm \frac{\sqrt{6-\lambda^2}}{\sqrt{6}}, 1 , 0\rp $ & $0$ & $  \lambda < \sqrt{6}$ & $ \frac{\lambda^2}{3}-1$ \\[8pt]
%&&&\\[-8pt]
 &  & (For $  \lambda = \sqrt{6}$, $\bar Q_{\phi}= \bar Q_{\rm kin}^+$)  & \\[8pt]
\hline
&&&\\[-8pt]
$\bar Q_{m\,\phi}= $ $\lp \frac{1}{\lambda} \sqrt{\frac{3}{2}},\pm\frac{1}{\lambda} \sqrt{\frac{3}{2}},1,0\rp $ & $1-\frac{3}{\lambda^2}$ & $  \lambda > \sqrt{3}$  & $0$\\[8pt]
%&&&\\[-8pt]
 &  & (For $  \lambda = \sqrt{3}$, $\bar Q_{m\phi}= \bar Q_{\phi}$)  & \\[8pt]
\hline
&&&\\[-8pt]
$\bar Q_{r}= $ $(0,0,1,\pm 1)$ & $0$ & $ \forall\, \lambda$ & $\frac{1}{3}$ \\[3pt]
\hline
&&&\\[-8pt]
$\bar Q_{r\,\phi}= $ $\left( \frac{1}{\lambda} \sqrt{\frac{8}{3}},\pm \frac{2}{\lambda\, \sqrt{3}},1, \pm \sqrt{1-\frac{4}{\lambda^2}} \right)$ & $0$ & $\lambda > 2$ & $\frac{1}{3}$ \\[8pt]
%&&&\\[-8pt]
 &  & (For $  \lambda = 2$, $\bar Q_{r\phi}= \bar Q_{\phi}$)  & \\[8pt]
\hline
\end{tabular}
\end{center}
\caption {Fixed points for the system \eqref{eq:system2} with the constraint \eqref{eq:Omegamxyu} and the exponential potential \eqref{potexp}. See the main text for their description.}
\label{tab:1}
\end{table}

Let us now discuss  the properties and stability of the  fixed points in the closed case. 
%and compare with the properties of the open case (see Appendix \ref{app1}).

\begin{itemize}
    \item $\bar Q_{{\rm kin}}^\pm$ -- {\sl kinetic domination}: the energy density is dominated by the kinetic energy of the scalar field, with $\bar x^2=1$, and thus $w_{\rm eff} =1$. These points are the only fully unstable fixed points, and thus all cosmological trajectories originate near them in the far past.  This is analogous to the open case.

    \item $\bar Q_m$ -- {\sl matter domination}: 
the energy density is dominated by matter, $\bar \Omega_m=1$, and thus $w_{\rm eff} =0$. This point is a saddle  and thus  cosmological trajectories may pass through it depending on the initial conditions. This is analogous to the point  $P_m$ in the open case (see table \ref{tab:3}).  

    \item $\bar Q_{k\,\phi}$ -- {\sl curvature scaling}: at this point the universe evolves under the influence of both curvature and the scalar field, but the expansion mimics curvature domination with $w_{\rm eff}=-\frac13$. This is analogous to the $P_{k \,\phi}$ point in the open case (see table \ref{tab:3}), however, the existence condition is the opposite: $\lambda< \sqrt{2}$. Contrary to the open case, this point is a saddle in the closed case. 

    \item $\bar Q_{\phi}$ -- {\sl scalar domination}: the energy density is dominated by the scalar, with $w_{\rm eff}= w_\phi$ and $\bar\Omega_\phi=1$. This is analogous to the point  $P_\phi$ in the open case and the stability properties are the same: stable for $\lambda\leq \sqrt{2}$, saddle otherwise.  For $\lambda < \sqrt{2}$ it therefore represents the late-time accelerating attractor for all cosmological trajectories. 

    \item $\bar Q_{m\,\phi}$ -- {\sl matter scaling}: the evolution is driven by the scalar and matter energy densities, but mimics a matter dominated evolution, with  $w_{\rm eff}=0$ and $\bar\Omega_m =1-3/\lambda^2$. This is analogous to the point $P_{m\,\phi}$ in the open case and it is a saddle.

    \item $\bar Q_{r}$ -- {\sl radiation domination}: The energy density is dominated entirely by radiation with  $w_{\rm eff} =1/3$. This is analogous to the point $P_r$ in the open case and it is a saddle.

    \item $\bar Q_{r\,\phi}$ -- {\sl radiation scaling}: the evolution is driven by both the scalar and radiation, though the evolution mimics a radiation dominated universe with $w_{\rm eff}=\frac13$, while $\bar\Omega_r = \bar u^2 =1-4/\lambda^2$. This is a saddle point and it is analogous to the point $P_{r\phi}$ in the open case.

\end{itemize}

In summary, there are three main differences with the open universe case (see Tables \ref{tab:3} and \ref{tab:4} in Appendix \ref{app1}). The first is that in the closed case there is no  curvature dominated point, as expected (in the open case, this is $P_k$ in table \ref{tab:4}). The second is that the curvature scaling point is not stable, and its existence condition is complementary to the open case: $\lambda< \sqrt{2}$ vs.~$\lambda> \sqrt{2}$, for the closed and open case respectively. Finally, whilst both the open and closed universe systems have a scalar dominated fixed point attractor when $\lambda < \sqrt{2}$, when $\lambda>\sqrt{2}$ the open universe has a curvature scaling fixed point attractor but the closed universe system has no attractor.  In other words, for the closed universe, the entire cosmological trajectory depends on the initial conditions.

\begin{figure}[H]
    \centering
    \includegraphics[width=0.5\linewidth]{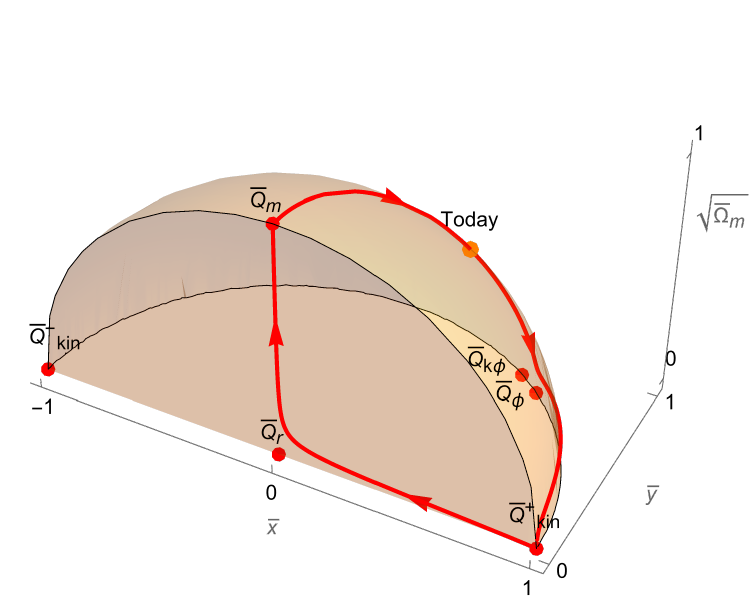}\includegraphics[width=0.5\linewidth]{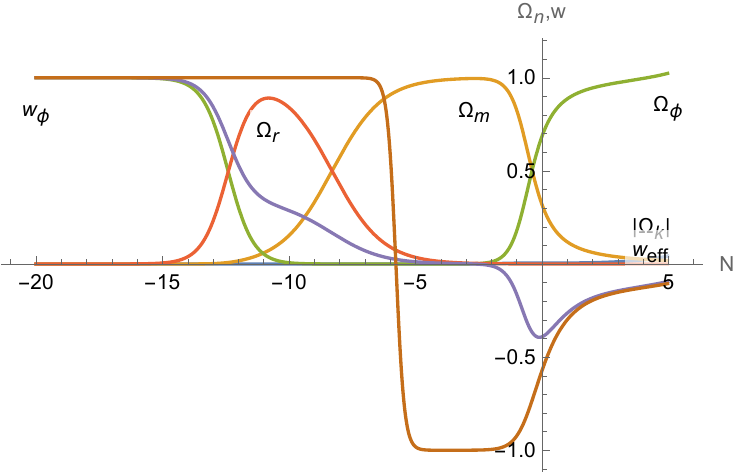}
    \caption{{\it Left:} a slice of the phase space for the dynamical system including quintessence, matter, radiation and positive spatial curvature, with the selection of variables $\bar{x}$, $\bar{y}$ and $\sqrt{\bar{\Omega}_m}$, for $\lambda=\sqrt{{8}/{3}}$.  Plotted in red is the trajectory determined by the initial conditions set today by $\Omega_{k,0}=-0.005000423845$, $\Omega_{\phi,0}=0.684587702938$, $\Omega_{r,0}=0.000080007654$ and $w_{\phi,0}=-0.573508930813358$, with: $\bar{z}_0 = {1}/{\sqrt{1-\Omega_{k,0}}}$, $\bar{x}_0=\sqrt{\Omega_{\phi,0}\bar{z}_0^{\,2} (1+w_{\phi,0})/2}$, $\bar{y}_0=\sqrt{\Omega_{\phi,0}\bar{z}_0^{\,2} - \bar{x}_0^{\,2}}$, $\bar{u}_0 = \sqrt{\Omega_{r,0}} \bar{z}_0$.  These values are taken from the run of CAMB described in the following section, with a large amount of precision necessary to ensure past radiation domination.  {\it Right:}  The corresponding evolution of density parameters, $\Omega_n$ ($\Omega_\phi$ in green, $\Omega_r$ in red, $\Omega_m$ in yellow, and $\Omega_k$ in blue) and equations of state ($w_\phi$ in brown and $w_{\rm eff}$ in purple), with respect to $N$, related to $\bar{N}$ as $dN = \bar{z} d\bar{N}$.  See main text for more details. }
    \label{fig:dynsys}
\end{figure}

In figure \ref{fig:dynsys} we present an example slice of the dynamical systems phase space for $\lambda=\sqrt{{8}/{3}}$, choosing to present the variables $\bar{x}, \bar{y}$ and $\sqrt{\bar{\Omega}_m}$ and dropping $\bar{u}$ and $\bar{z}$ (but recall the constraint $\bar{\Omega}_m = 1 -\bar{x}^2-\bar{y}^2-\bar{u}^2$).  We also plot in the phase space an example trajectory, with initial conditions fixed by present-day values for the density parameters and $w_\phi$, chosen to match  the CAMB runs described in the following subsection.  We learn that the cosmological solution starts in a kination epoch, from the fixed point $\bar{Q}^+_{\rm kin}$, passes through radiation and matter domination as it approaches $\bar{Q}_m$ and $\bar{Q}_r$, respectively, meets the point corresponding to today (as it must), where there is transient acceleration, and eventually returns to $\bar{Q}^+_{\rm kin}$, another kination epoch. Recall that the latter fixed point is a saddle, thus we do not expect it to be a final destination. See \cite{Andriot:2024jsh} for similar phase space diagrams for the open universe.

%\bigskip

\begin{table}[H]
\begin{center}
\centering
\begin{tabular}{| l | c | c | c |}
\hline
\cellcolor[gray]{0.9} & \cellcolor[gray]{0.9} & \cellcolor[gray]{0.9} &  \cellcolor[gray]{0.9} \\[-8pt]
\cellcolor[gray]{0.9} Point & \cellcolor[gray]{0.9} Eigenvalues &  \cellcolor[gray]{0.9} Stability &  \cellcolor[gray]{0.9} Existence \\[5pt]
\hline
&&&\\[-10pt]
&& $\bar Q_{{\rm kin}}^-:$ Fully unstable \phantom{for $\lambda \leq \sqrt{6}$} & \\
$\bar Q_{{\rm kin}}^\pm $ & $ \lp 4,3,3 \mp \lambda \sqrt{\frac32}, 1 \rp$  & $\bar Q_{\rm kin}^+:$ Fully unstable for $\lambda \leq \sqrt{6}$ & - \\
&& $\bar Q_{{\rm kin}}^+:$ Saddle for $\lambda >\sqrt{6}$ \phantom{Fully i} &\\[6pt]
\hline
&&&\\[-8pt]
$\bar Q_{k} $ & $(-\frac{3}{2},\frac{3}{2},-\frac{1}{2},1)$  & Saddle & - \\[4pt]
\hline
&&&\\[-10pt]
$\bar Q_{k\,\phi}$ & $\lp -\frac{\lambda}{\sqrt{2}}, -\frac{\lambda}{\sqrt{2}}, \frac{-\lambda +\sqrt{8-3\lambda^2}}{\sqrt{2}},-\frac{\lambda +\sqrt{8-3\lambda^2}}{\sqrt{2}} \rp $  & Saddle & $\lambda < \sqrt{2}$ \\[6pt]
\hline
&&&\\[-7pt]
 &  & Stable for $\lambda < \sqrt{2}$ & \\[-8pt]
$\bar Q_\phi$ & $ \lp \frac{\lambda^2}{2}-3,\frac{\lambda^2}{2}-3 ,\lambda^2-2 , \lambda^2-2\rp$ && $\lambda < \sqrt{6}$ \\[-10pt]
&& Saddle for $\lambda > \sqrt{2}$ & \\[6pt]
\hline
&&&\\[-10pt]
$\bar Q_{m\,\phi} $ & $\lp -\frac12,1, \frac{3(-\lambda+\sqrt{24-7\lambda^2})}{4\lambda},-\frac{3(\lambda +\sqrt{24-7\lambda^2})}{4\lambda}, -\frac12 \rp $  & Saddle & $\lambda > \sqrt{3} $ \\[6pt]
\hline
&&&\\[-8pt]
$\bar Q_{r} $ & $ (2,2,1,-1)$  & Saddle & - \\[5pt]
\hline
&&&\\[-8pt]
$\bar Q_{r\,\phi} $ & $ \lp 1,2,\frac{-\lambda+\sqrt{64-15 \lambda^2}}{2\lambda} , -\frac{\lambda+\sqrt{64-15 \lambda^2}}{2\lambda} \rp$  & Saddle & $\lambda > 2 $ \\[5pt]
\hline
\end{tabular}
\end{center}
\caption {Summary of the stability analysis for the fixed points in table \ref{tab:2} corresponding to  a closed universe. }
\label{tab:2}
\end{table}

\subsection{Background  evolution}\label{sec:backevol}
We now study  the cosmological evolution of the curved exponential quintessence.  We implement the curved quintessence model in the cosmological Boltzmann code \texttt{CAMB}~\cite{Lewis:1999bs,Howlett:2012mh}. 
 We choose initial conditions for the field: $\phi_i = 0,\, \dot{\phi}_i=0$, deep in the radiation era.\footnote{Note that a non-zero value of $\phi_i$ can be absorbed into a redefinition of $V_0$. Moreover,  any non-zero initial velocity is quickly washed out due to Hubble friction.}   For a given input value of $\lambda$ and other background density parameters \mbox{$\{\Omega_{\rm b},\Omega_{\rm c},\Omega_{k},\Omega_{\rm r}\}$}, the amplitude of the potential $V_0$ is tuned by the code and adjusted to obtain the correct $\Omega_{\phi}$ today, in order  to satisfy Eq.~\eqref{eq:friO}.  
 
 We consider two cases  motivated by string theory set-ups as discussed in \cite{Andriot:2024jsh}, $\lambda = \sqrt{3}, \sqrt{8/3}$, as well as a larger and smaller value motivated by our further cosmological analysis in the next section. The field typically remains frozen due to Hubble friction in the radiation era and starts to evolve much later when dark energy begins to dominate (see figures \ref{fig:ev_phi_Kneg}, \ref{fig:ev_phi_Kpos} below). 
We can understand the evolution from the dynamical system analysis. For  $\lambda=\sqrt{3}$, for both  open and closed universes, the matter scaling points merge with the scalar dominated point: $P_{m\phi} = P_\phi$ and  $\bar Q_{m\phi}=\bar Q_\phi$ respectively.   For the open case, there is a curvature scaling point $P_{k\phi}$, but not in the closed case. 
For $\lambda=\sqrt{8/3}$, for both closed and open universes, the matter scaling points do not exist, while the scalar dominated points still exist and are saddles ($P_{\phi}, \bar Q_\phi$). Further, the open curvature scaling point ($P_{k\phi}$) also exists (and is stable). 
Finally we also plot the evolution for $\lambda =1/2$ and  $\lambda =2$. For $\lambda=1/2$ the closed curvature scaling point exists, but not for the open case. The scalar dominated point instead exists for both open and closed cases, it gives rise to acceleration and it is stable. For $\lambda=2$,  the closed scaling point does not exist, but it does for the open case, while the scalar dominated point exists in both cases and it is a saddle.  For all the values of  $\lambda<\sqrt{2}$, the scalar fixed point is an accelerating attractor. In this case,  the universe will evolve towards an accelerating future with a cosmological horizon.

\subsubsection{Open universe}

\begin{figure}[H]
\centerline{
\includegraphics[width=1.1\textwidth]{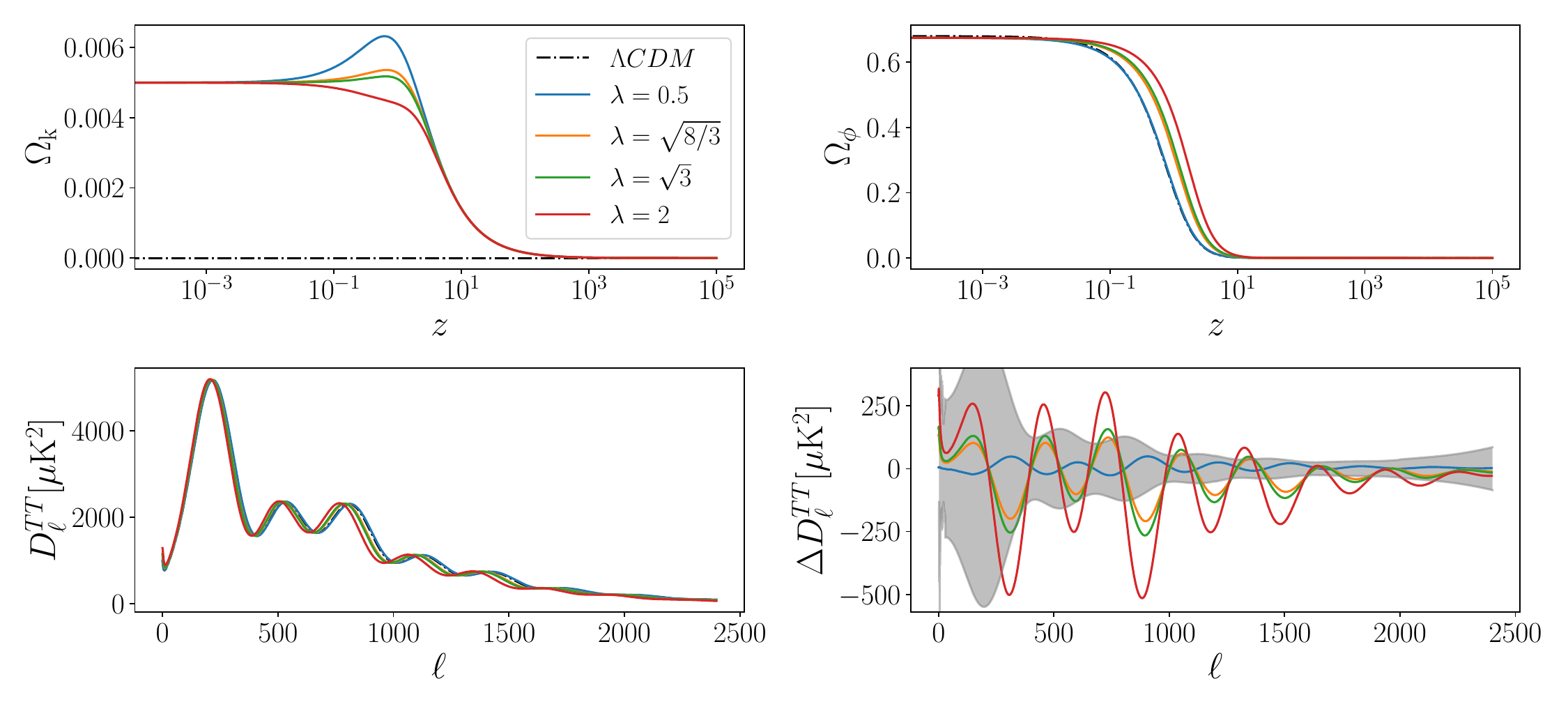}}
\caption{{\em Upper panel}: Curvature and scalar density parameter evolution for various values of $\lambda$ for an open universe $k<0$. {\em Lower panel}: angular power spectrum $D_{\ell}^{TT}$ and the residuals $\Delta D_{\ell}^{TT}$ with respect to $\Lambda$CDM , for the same values of $\lambda$ and $k<0$. The grey shaded regions represent the error bars on $D_{
\ell}^{TT}$ from \textit{Planck}~\cite{Planck:2019nip}.} 
\label{fig:Omegas_Kneg}
\end{figure}

In figure \ref{fig:Omegas_Kneg}  we show the evolution of the curvature and scalar   density parameters  for the  values of $\lambda$ described therein (at the top left panel). In the bottom left panel, we show the deviation of our model from $\Lambda$CDM by plotting the temperature angular power spectrum $D_{\ell}^{TT}\equiv \ell (\ell +1) C_{\ell}^{TT}/(2\pi)$, and we plot the difference $\Delta D_{\ell}^{TT} \equiv D_{\ell}^{TT,\Lambda \rm{CDM}} -D_{\ell}^{TT,q\rm{CDM}+\Omega _k}$ in the bottom right panel.

\begin{figure}[H]
\centerline{
\includegraphics[width=1.1\textwidth]{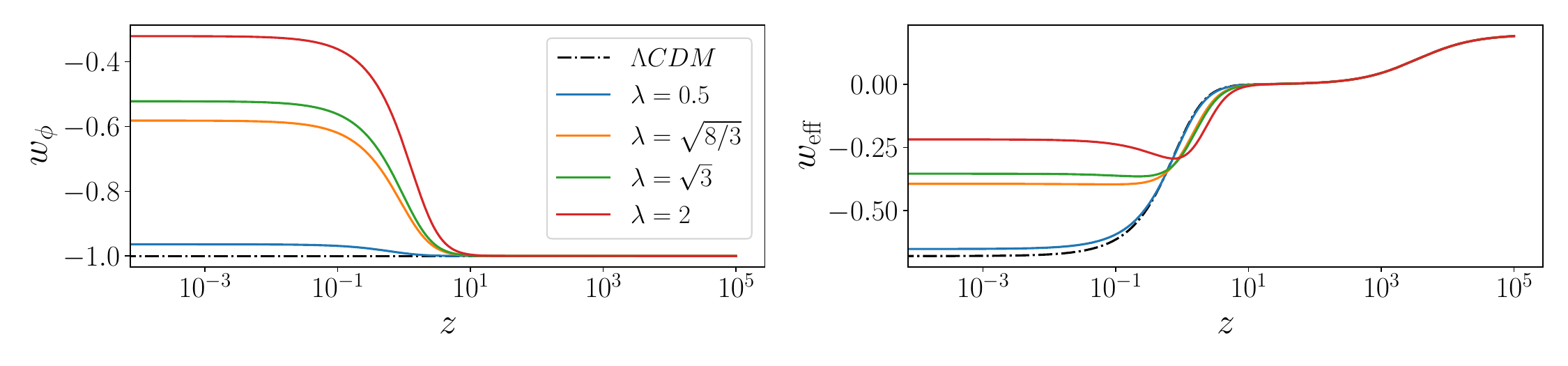}}
\caption{Equations of state evolution for the same parameter values as in figure \ref{fig:Omegas_Kneg}  computed from {\tt CAMB}. }
\label{fig:w_DE_w_eff_Kneg}
\end{figure}

In figure 
\ref{fig:w_DE_w_eff_Kneg} we show the evolution of the corresponding scalar and effective equations of state, $w_{\phi}$ and $w_{\rm eff}$. The values of  the parameters today together with $V_0$, as computed by {\tt CAMB}, are given in table \ref{energy_densities_kn}.

\begin{figure}[H]
\centerline{
\includegraphics[width=0.5\textwidth]{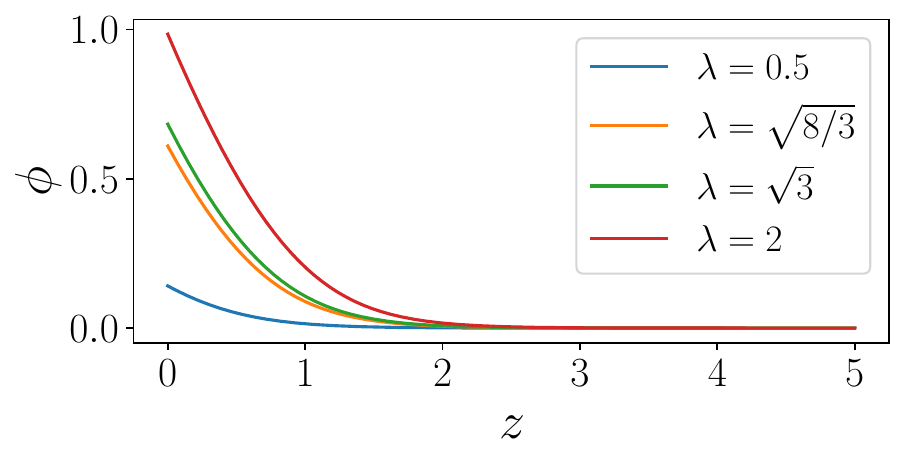}}
\caption{Scalar field evolution for the open quintessence case.}
\label{fig:ev_phi_Kneg}
\end{figure}

\renewcommand{\arraystretch}{1.5}
\begin{table}[H]
\begin{center}
\centering
\begin{tabular}{| c | c | c |}
\hline
\rowcolor{gray!30} 
$\lambda$ & $V_0 \,\,({\rm eV}^2)$  & $w_{\phi,0}$\\
\hline
 $0.5$ & $1.0715\times 10^{-7}$  & $-0.96369$\\
\hline
 $\sqrt{8/3}$ & $2.1765\times 10^{-7}$  & $-0.58169$\\
\hline
 $\sqrt{3}$ & $2.5255\times 10^{-7}$ & $-0.52205$\\
\hline
 $2$ & $4.8067\times 10^{-7}$ & $-0.32093$\\
\hline
\end{tabular}
\end{center} 
\caption{Parameter values  for $k < 0$ for the plots in Figures \ref{fig:Omegas_Kneg},  \ref{fig:w_DE_w_eff_Kneg} and \ref{fig:ev_phi_Kneg} as
computed from {\tt CAMB}.   The associated values for today's density parameters are $\Omega_{k,0}=0.00500$, $\Omega_{c,0}=0.27018$, $\Omega_{b,0}=0.04872$, and $\Omega_{\phi,0}=0.67459$.}
\label{energy_densities_kn}
\end{table}

\subsubsection{Closed universe}

\begin{figure}[H]
\centerline{
\includegraphics[width=1\textwidth]{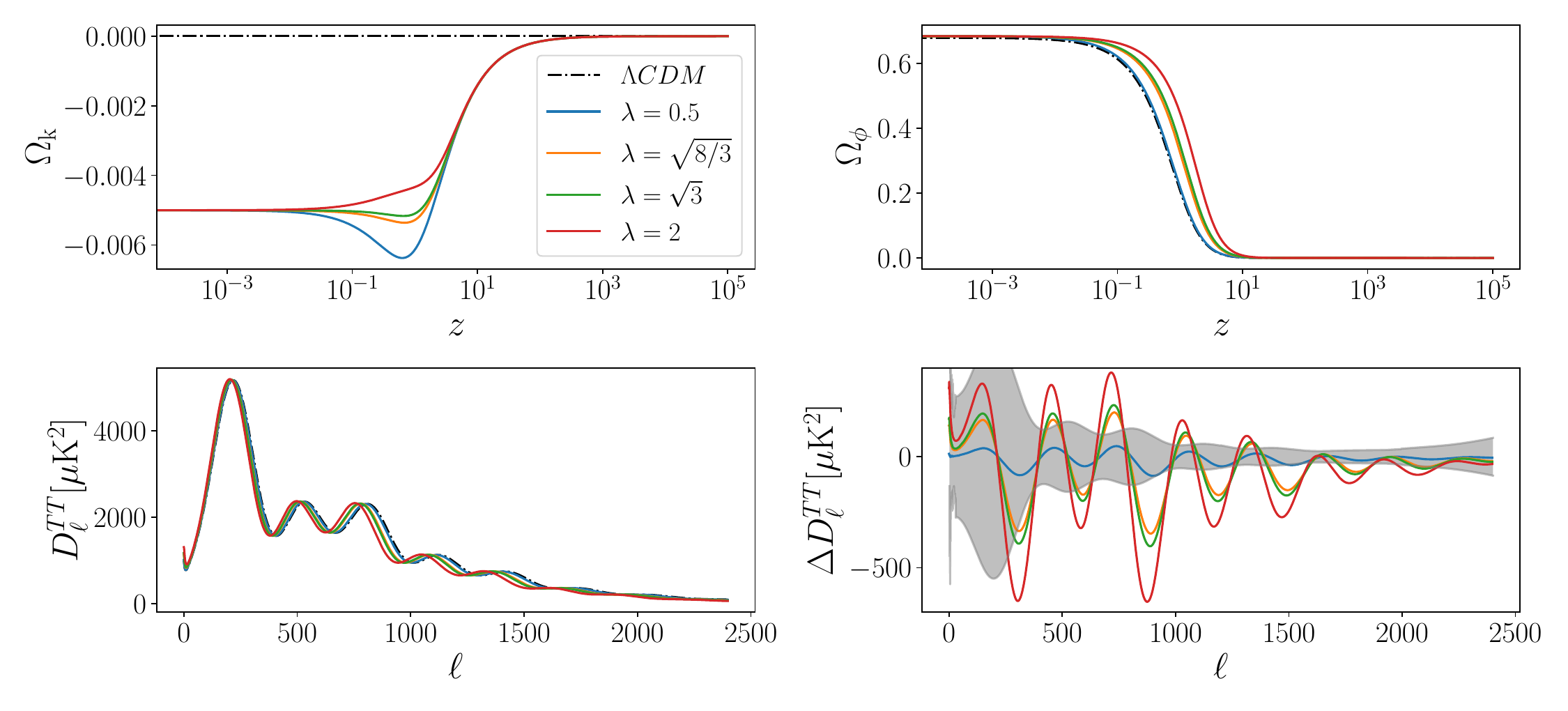}}
\caption{Curvature and scalar density parameter  evolution for a closed universe, $k>0$ for various values of $\lambda$ (upper panel). Angular power spectrum $D_{\ell}^{TT}$ and residuals $\Delta D_{\ell}^{TT}$
for same values of $\lambda$ and $k>0$ (lower panel). The grey shaded regions represent the error bars on $D_{
\ell}^{TT}$ from \textit{Planck}~\cite{Planck:2019nip}. }
\label{fig:Omegas_Kpos}
\end{figure}

\begin{figure}[H]
\centerline{
\includegraphics[width=1.1\textwidth]{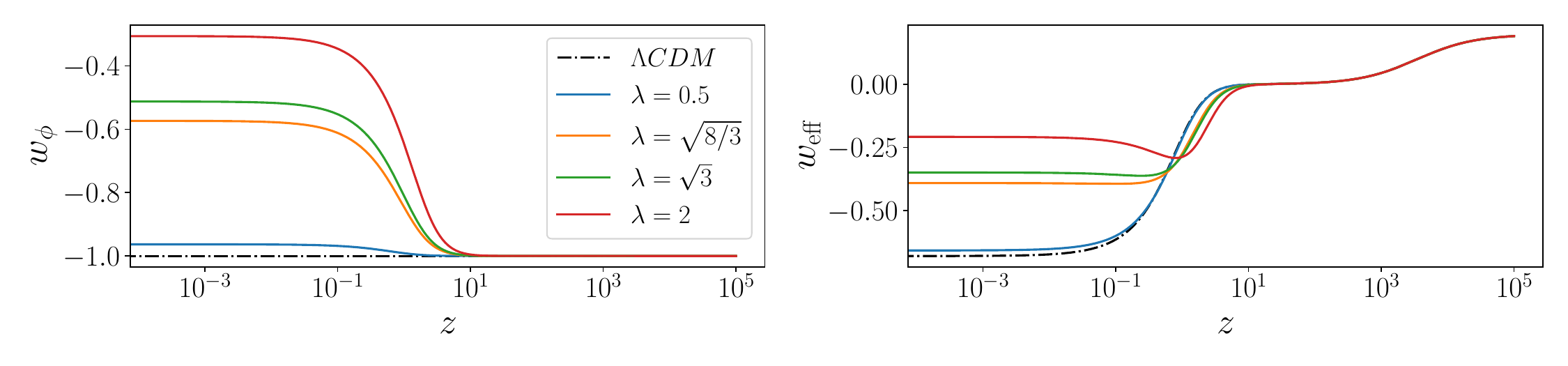}}
\caption{Evolution of the equations of state for $k>0$ for the different values of $\lambda$ discussed in the text.}
\label{fig:w_DE_w_eff_Kpos}
\end{figure}

\begin{figure}[H]
\centerline{
\includegraphics[width=0.5\textwidth]{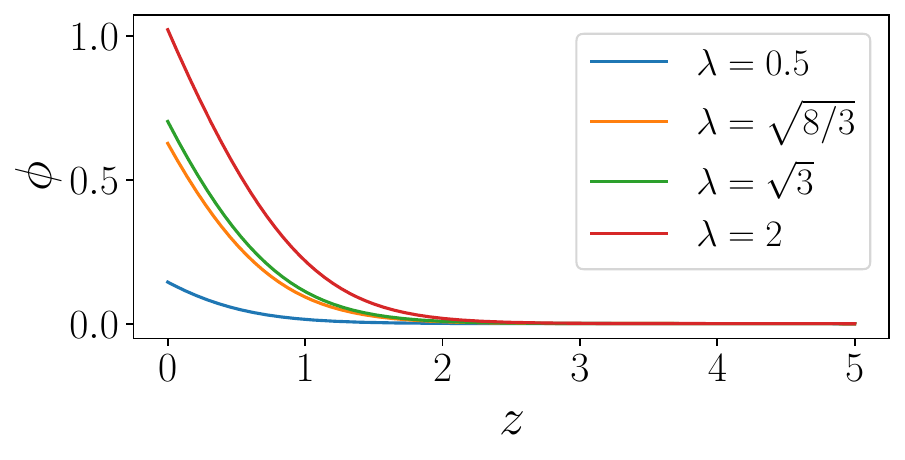}}
\caption{Scalar field evolution for the closed quintessence case. }
\label{fig:ev_phi_Kpos}
\end{figure}

In figure \ref{fig:Omegas_Kpos}  we show the evolution of the density parameters 
as well as the angular power spectrum $D_{\ell}^{TT}$ and  $\Delta D_{\ell}^{TT}$, for the  values of $\lambda$ described above. In figure 
\ref{fig:w_DE_w_eff_Kpos} 
we show evolution of the scalar and effective equations of state and in 
figure \ref{fig:ev_phi_Kpos}
we show the corresponding evolution of the scalar field. The values of current-day   parameters as computed from {\tt CAMB} for these cases are given in table \ref{energy_densities_kp}.

% CLosed case 
%%%%%%%%%%%

\renewcommand{\arraystretch}{1.5}
\begin{table}[H]
\begin{center}
\centering
\begin{tabular}{| c | c | c  |}
\hline
\rowcolor{gray!30} 
$\lambda$ & $V_0\,\,({\rm eV}^2)$ & $w_{\phi,0}$\\
\hline
 $0.5$ & $1.0886\times 10^{-7}$  & $-0.96300$\\
\hline
 $\sqrt{8/3}$ & $2.2576\times 10^{-7}$ & $-0.57350$\\
\hline
 $\sqrt{3}$ & $2.6350\times 10^{-7}$  & $-0.51259$\\
\hline
 $2$ & $5.1973\times 10^{-7}$ & $-0.30628$\\
\hline
\end{tabular}
\end{center} 
\caption{Parameter values  as computed from {\tt CAMB} for the plots in figures \ref{fig:Omegas_Kpos}, \ref{fig:w_DE_w_eff_Kpos} and  \ref{fig:ev_phi_Kpos} ($k>0$).  The associated values for today's density parameters are $\Omega_{k,0}=-0.00500$, $\Omega_{c,0}=0.27018$, $\Omega_{b,0}=0.04872$, and $\Omega_{\phi,0}=0.68459$. }
\label{energy_densities_kp}
\end{table}

Comparing the open and closed universes,
 we  learn that  the closed universe yields slightly higher values for the dark energy density parameter today, $\Omega_{\phi}^0$, for a given $\lambda$.  This is to be expected from \eqref{eq:friO}.  On the other hand, the value of $w_{\phi}$ for a given $\lambda$ is slightly less negative in the closed case (see \cref{energy_densities_kn} and \cref{energy_densities_kp}).

%%%%%%%
\section{Cosmological constraints}\label{sec:constrains}

With the modified \texttt{CAMB} code, we are now ready to analyse our  curved quintessence model against cosmological datasets. We explore the parameter space of the model using Markov Chain Monte-Carlo (MCMC) methods, varying $\lambda$ alongside the other cosmological parameters \mbox{$\{\Omega_{\rm b}h^2,\Omega_{\rm c}h^2,\Omega_k,H_0,\tau,A_s,n_s\}$}.\footnote{The parameter $\lambda$ is the only free parameter of the quintessence model since $V_0$ is getting tuned at each step of the MCMC to satisfy the budget equation~\eqref{eq:friO}.} We use the following cosmological likelihoods: %{\st{datasets:}}
 \begin{enumerate}
    \item{CMB from \textit{Planck}:}
\begin{itemize}
    \item \textit{Planck} 2018 low-$\ell$ temperature and polarisation likelihood~\cite{Aghanim:2019ame}.
    \item  \textit{Planck} high-$\ell$ CamSpec TTTEEE temperature and polarization likelihood using \texttt{NPIPE} (\textit{Planck} PR4) data~\cite{Rosenberg:2022sdy}.
    \item \textit{Planck} 2018 lensing likelihood~\cite{Aghanim:2018oex}.
\end{itemize}
 In what follows, we collectively denote all the \textit{Planck} likelihoods as `CMB'.
    \item BAO likelihoods from DESI DR1~\cite{DESI:2024lzq,DESI:2024mwx,DESI:2024uvr} consisting of bright galaxy survey (BGS),
    luminous red galaxies (LRG), emission line galaxies (ELG), quasars 
    and Lyman-$\alpha$ (Ly$\alpha$) data which cover a total redshift range \mbox{$0.1<z<4.2$}. Sloan Digital Sky Survey (SDSS) DR16 likelihoods~\cite{Alam:2020sor} covering a total redshift range \mbox{$0.07<z<3.5$} for the ELG, LRG, quasar and Ly$\alpha$ tracers.
     \item  Pantheon+~\cite{Brout:2022vxf}, Union3~\cite{Rubin:2023ovl} and DES-Y5~\cite{DES:2024tys} type Ia supernovae samples. Altogether, these samples consist of about $4000$ supernovae which cover a redshift range $ 0.001 < z <2.26$. Note that the samples cannot be combined since Pantheon+ and Union3 have 1363 supernovae in common while DESY5 has 194 low redshift supernovae shared with the other two.

\end{enumerate}

We implement wide uniform priors on all the cosmological parameters. The likelihoods are sampled using the \texttt{MCMC} sampler~\cite{Lewis:2002ah,Lewis:2013hha}, through its interface with \texttt{Cobaya}~\cite{Torrado:2020dgo}. We continue the sampling until we reach a value $R-1=0.03$ for the Gelman-Rubin diagnostic. The resulting chains are analysed and plotted with~\texttt{GetDist} package~\cite{Lewis:2019xzd}. At the end of the sampling, we also run the \texttt{Py-BOBYQA}~\cite{Bobyqa1,Bobyqa2} minimizer through the \texttt{Cobaya} interface to find the best-fit point and the corresponding $\chi^2$ values.

\begin{figure}
    \centering
    \includegraphics[width=0.75\linewidth]{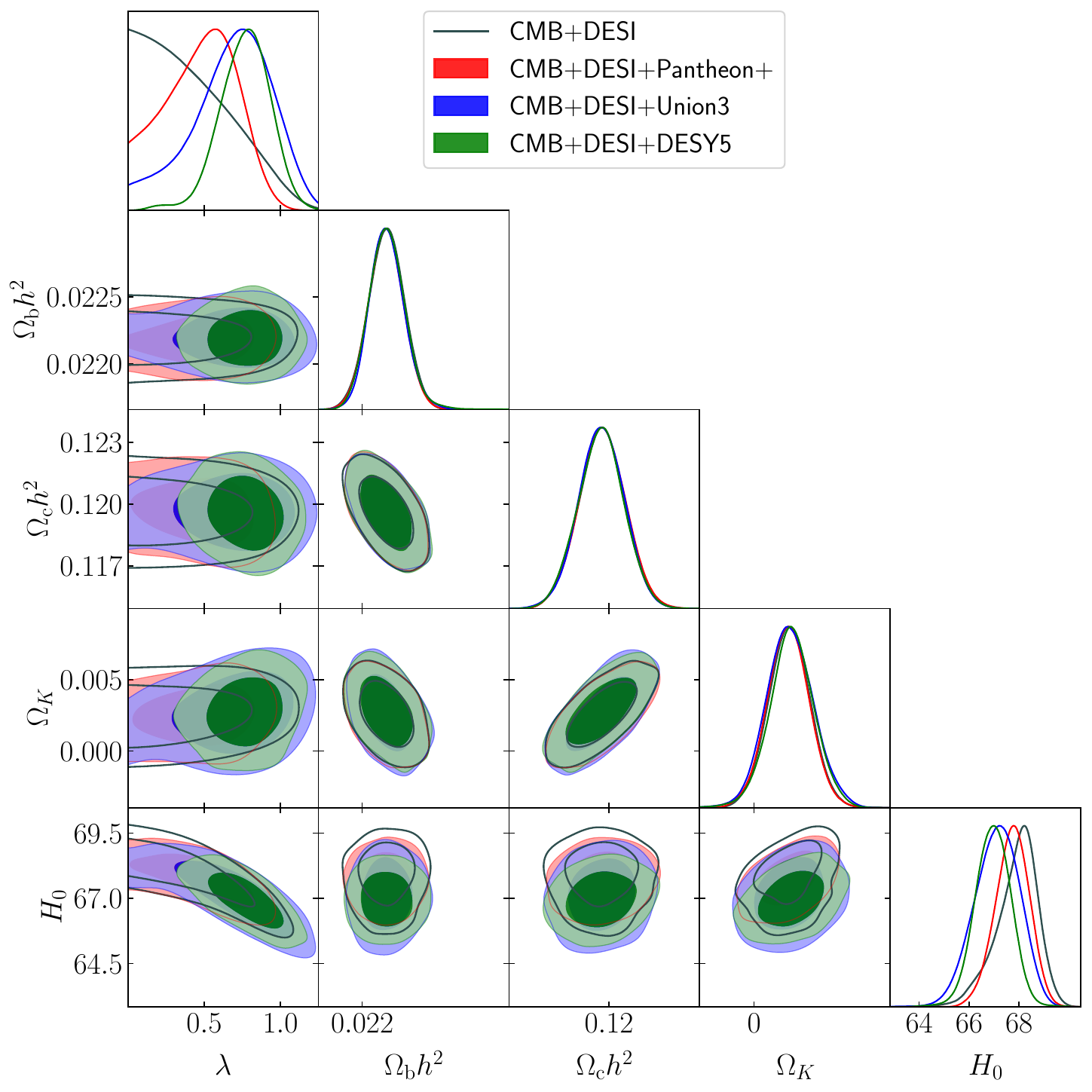}
    \caption{Constraints on cosmological parameters from the combinations of datasets described in the main text. See \cref{app:full_constraints} for the inclusion of all the other parameters.}
\label{fig:contours_5param_sn}
\end{figure}

The recent DESI BAO data alone, as well as in conjunction with the supernovae datasets, have already been shown to have a preference for a time varying dark energy equation of state as compared to $\Lambda$CDM~\cite{DESI:2024mwx,Calderon:2024uwn,Lodha:2024upq}. When the same datasets are used with our exponential quintessence model as the underlying dark energy model, this manifests as a slight preference for a non-zero $\lambda$, resulting in the following marginalised mean and limits (at $68\%$ C.L) shown in \cref{tab:param_limits_sn}.

\begin{table}[H]
    \centering
\begin{tabular} { |c| c| c| c| c|}

\hline
\rowcolor{gray!30} 
 \bf{Parameter} &  {\bf CMB+DESI} & {\bf +Pantheon+} &  {\bf +Union3+} &  {\bf +DESY5} \\
\hline
{\boldmath$\lambda        $} & $< 0.537                   $ & $0.48^{+0.28}_{-0.21}      $ & $0.68^{+0.31}_{-0.20}      $ & $0.77^{+0.18}_{-0.15}$\\

{\boldmath$\Omega_k       $} & $0.0026\pm 0.0015          $ & $0.0025\pm 0.0015          $ & $0.0028^{+0.0016}_{-0.0019}$ & $0.0027\pm 0.0016          $\\

{\boldmath$\Omega_\mathrm{c} h^2$} & $0.1196\pm 0.0012          $ & $0.1197\pm 0.0012          $ & $0.1195\pm 0.0012          $ & $0.1195\pm 0.0012          $\\

{\boldmath$H_0            $} & $67.89^{+0.96}_{-0.61}     $ & $67.73^{+0.72}_{-0.64}     $ & $67.12^{+0.97}_{-0.83}            $ & $66.95\pm 0.72            $\\

{\boldmath$\Omega_\mathrm{b} h^2$} & $0.02219\pm 0.00014        $ & $0.02219\pm 0.00013        $ & $0.02220^{+0.00013}_{-0.00015}$ & $0.02221\pm 0.00013        $\\
\hline
\end{tabular}
    \caption{Parameter means and $68\%$ limits for the combination of the CMB+DESI and with the addition of the different supernovae datasets to the CMB+DESI baseline.}
    \label{tab:param_limits_sn}
\end{table}
The marginalised 1D and 2D joint distributions for these parameters are plotted in figure \ref{fig:contours_5param_sn}. As we can see, the combination of CMB and DESI BAO data is not able to significantly constrain the quintessence model parameter $\lambda$. However, this changes with the addition of the supernovae datasets and 
depending on the dataset chosen, the obtained values for $\lambda$ lie about $2\text{--}4\sigma$ away from the $\lambda=0$ case that corresponds to a cosmological constant. In~\cref{app:full_constraints}, we plot the contours for the full parameter set, including $\{n_s,A_s,\tau\}$. There,  we also present the results of the MCMC analysis for the dataset combination CMB+DESI+Pantheon+, fixing $\omk=0$ and compare it to the free spatial curvature case~(\cref{fig:curved_flat}). We do not find any major difference for the cosmological parameters constraints, in particular for $\lambda$, obtaining $\lambda= 0.42\pm 0.22$ at 68\% C.L.

\begin{figure}[h]
    \centering
    \includegraphics[width=0.75\linewidth]{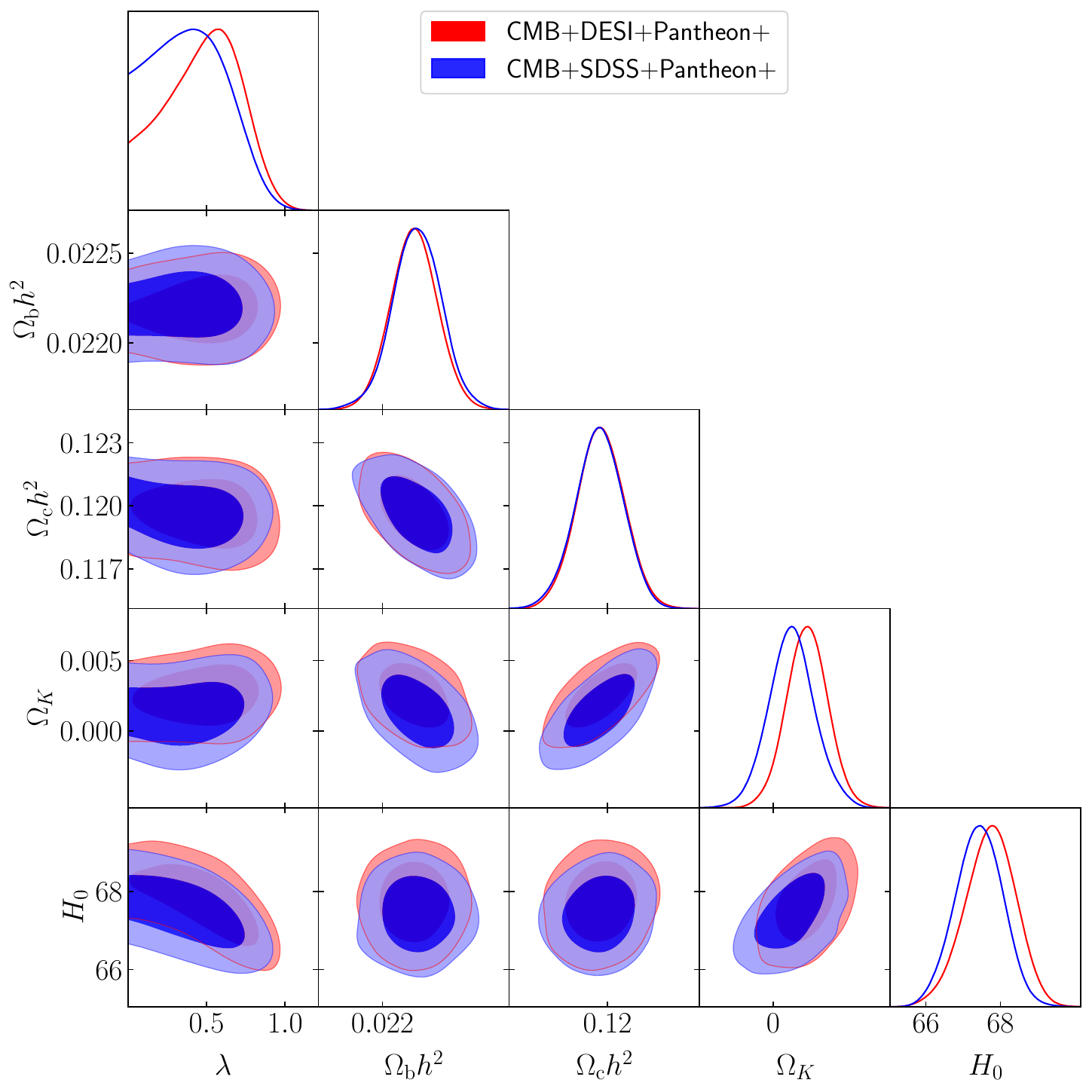}
    \caption{Constraints on cosmological parameters for CMB+DESI+Pantheon+ and CMB+SDSS+Pantheon+. See \cref{app:full_constraints} for the inclusion of all the other parameters.}
    \label{fig:5param_diffbao}
\end{figure}

Fixing our supernovae dataset to Pantheon+, we also assess the effect of replacing the full DESI BAO data with the SDSS data, finding nearly identical results for most cosmological parameters with a slightly lower deviation from a cosmological constant in terms of the preferred values of $\lambda$ (see \cref{fig:5param_diffbao} and \cref{tab:5param_diffbao}). The result could be attributed to the increased constraining power of the DESI data as compared to SDSS. For all the datasets used in this section, we also observe a mild shift towards $\Omega_k>0$ with the difference from $\Omega_k=0$ being less than~ $2\sigma$.  There is no noticeable shift for any of the other cosmological parameters. 

\begin{table}[H]
    \centering
\begin{tabular} { |c |c|c|}

\hline
\rowcolor{gray!30} 
\bf{Parameter} & {\bf CMB+DESI+Pantheon+} & {\bf CMB+SDSS+Pantheon+} \\
\hline
{\boldmath$\lambda        $} & $0.48^{+0.28}_{-0.21}      $ & $0.40^{+0.20}_{-0.29}      $\\

{\boldmath$\Omega_k      $} & $0.0025\pm 0.0015          $ & $0.0014\pm 0.0017        $\\

{\boldmath$\Omega_\mathrm{c} h^2$} & $0.1197\pm 0.0012          $ & $0.1197^{+0.0013}_{-0.0012}$\\

{\boldmath$H_0            $} & $67.73^{+0.72}_{-0.64}     $ & $67.44\pm 0.64     $\\

{\boldmath$\Omega_\mathrm{b} h^2$} & $0.02219\pm 0.00013        $ & $0.02220^{+0.00013}_{-0.00016}$\\
\hline
\end{tabular}
\caption{Parameter means and 68\% limits.}
    \label{tab:5param_diffbao}
\end{table}
Considering that the new DESI dataset favours dynamical dark energy, we can compare how well our model fits the data against the CPL parametrisation presented in~\cite{DESI:2024mwx}. For this, we first compute the $\Delta\chi^2$ for both the $q$CDM+$\omk$ model and the $w_0w_a$CDM+$\Omega_k$ model with respect to 
$\Lambda$CDM +$\omk$, i.e. 
\be
\Delta\chi^2_{\rm model} \equiv \chi^2_{\rm model} 
- \chi^2_{\rm \Lambda CDM}\,.
\ee
where the $\chi^2=-2\ln \mathcal{L}_{\rm max}$ for a given model.
We find for the CMB+DESI+Pantheon+ combination
\be
\Delta\chi^2_{q \rm CDM } = -1.8 \qquad \text{and} \qquad  \Delta\chi^2_{w_0 w_a \rm CDM} = -6\,,
\ee
showing that both models provide a better fit to the data compared to $\Lambda$CDM$+\Omega_k$. This is to be expected since we can always recover $\Lambda$CDM from the two models by fixing, respectively, $\lambda=0$ or $w_0=-1,\,w_a=0$. We remind the reader that the $q$CDM model features one additional parameter compared to $\Lambda$CDM whereas the $w_0 w_a$ model has two.

When comparing cosmological models using their best-fit $\chi^2$ values, one should also take into account the number of free parameters of the model. For non-nested models,\footnote{It is only for nested models that the maximum (log) likelihood ratio or $\Delta\chi^2$ is approximately $\chi^2$ distributed~\cite{Liddle:2004nh}. The AIC has no such requirement but is a simplistic comparison method. A more rigorous comparison would be to compute the Bayesian evidences but this is computationally quite expensive.} a simple method to compare the quality of the fit to the data while at the same time accounting for the number of model parameters is provided by the Akaike information criterion (AIC)~\cite{AIC,Liddle:2004nh}. We compute this quantity for the $q$CDM$+\Omega_k$ and  $w_0 w_a$CDM$+\Omega_k$  models. The AIC value for a given model is defined as
\begin{align}
    \mathrm {AIC} \,=\,2n-2\ln \mathcal{L}_{\rm max}\,,
\end{align}
where $\mathcal{L}_{\rm max}$ denotes the maximum likelihood value for the model and $n$ is the number of free parameters (note $n_{w_0w_a\rm{CDM}}-n_{q\rm{CDM}}=(7+2)-(7+1)=1$). AIC  represents the information loss in using a particular model to represent the true underlying process with the best fitting model among the candidate models having the smallest AIC value, i.e. the lowest information loss~\cite{AIC,Liddle:2004nh}. 
For our analysis {with CMB+DESI+Pantheon+}, we find 
\begin{align}
    \mathrm {AIC}_{w_0 w_a\rm CDM} - \mathrm {AIC}_{q\rm CDM} = -2.2\,,\quad   \mathrm {AIC}_{q\rm CDM} - \mathrm {AIC}_{\Lambda\rm CDM} = 0.2\  %\qquad \text{(CMB+DESI+Pantheon+)},
\end{align}
which indicates a preference, albeit not very strong, for the $w_0 w_a$ parametrisation over the exponential quintessence model {and no preference between $q$CDM and $\Lambda$CDM}. On replacing the Pantheon+ dataset with Union3 or DESY5, the results change slightly e.g. \mbox{${\mathrm {AIC}}_{w_0 w_a\rm CDM} - {\mathrm {AIC}}_{q\rm CDM} = -3$} {and \mbox{${\mathrm {AIC}}_{q\rm CDM} - {\mathrm {AIC}}_{\Lambda\rm CDM} = -2.3$}} for CMB+DESI+Union3. {The increase in preference for $q$CDM over $\Lambda$CDM comes from the fact that these datasets require a larger deviation from $\Lambda$CDM compared to Pantheon+, as seen in the preferred values of $\lambda$ (and also with the CPL parametrisation~\cite{DESI:2024mwx}).}
{The preference for $w_0w_a$CDM over $q$CDM} is likely due to DESI + supernovae data~(as seen in~\cite{DESI:2024mwx,Calderon:2024uwn}) indicating a preference for phantom-like behaviour for dark energy ($w<-1$) in the past and going towards $w>-1$ near $z\approx 0$ more rapidly than what can be obtained in the $q$CDM model. 

Note however that the data does not constrain $w(z)$ directly but only indirectly through $H(z)$ and integrals of $H(z)$ that enter when calculating cosmological distances to a given redshift.  In particular, $w(z)$ is well constrained close to the point $z=0.4$ with the uncertainties increasing, especially at higher redshifts~\cite{DESI:2024mwx,Calderon:2024uwn}. Thus, to provide a good fit to the data a given quintessence model need not match the entire evolution of $w(z)$ with that in the $w_0 w_a$ model, but rather match $H(z)$ (or the distances) to a given accuracy, as pointed out in~\cite{Shlivko:2024llw}. According to   the latter's analysis based on this matching procedure, hilltop and plateau type models of quintessence may fit the data better as compared to the exponential model. Analysis of quintessence models from the DESI collaboration, based on parametrisations of classes of quintessence models (thawing, emergent, mirage) rather than using specific models, also finds that quintessence models can significantly improve the fit to the data w.r.t $\Lambda$CDM~\cite{Lodha:2024upq}, with the mirage class faring the best. 
\begin{figure}
    \centering
    \includegraphics[width=0.48\linewidth]{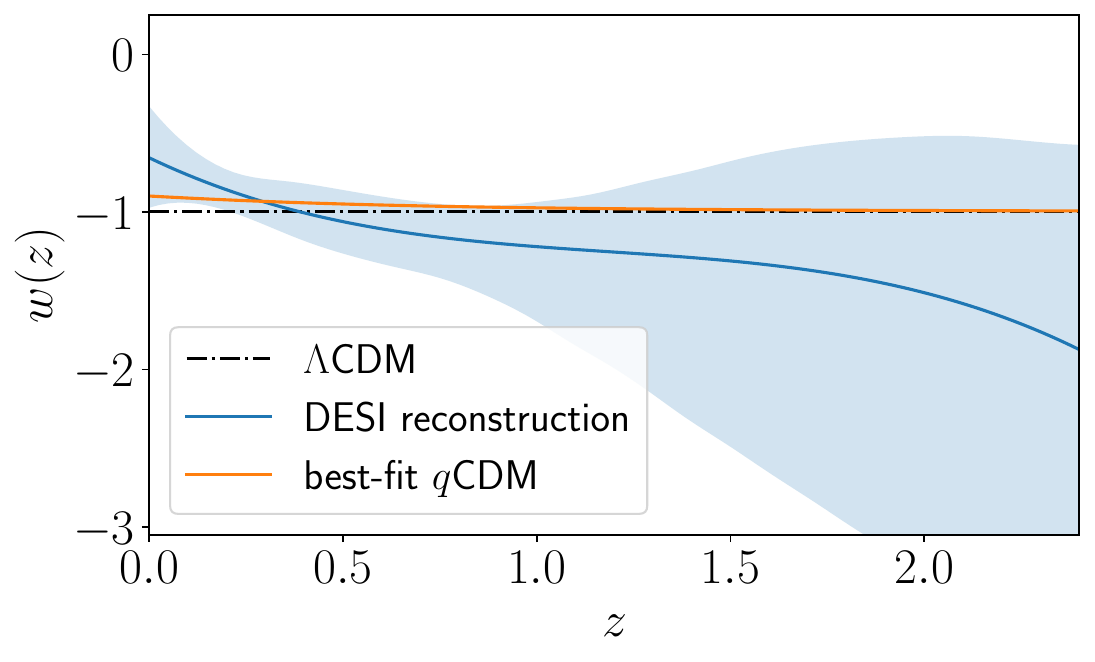}
    \includegraphics[width=0.48\linewidth]{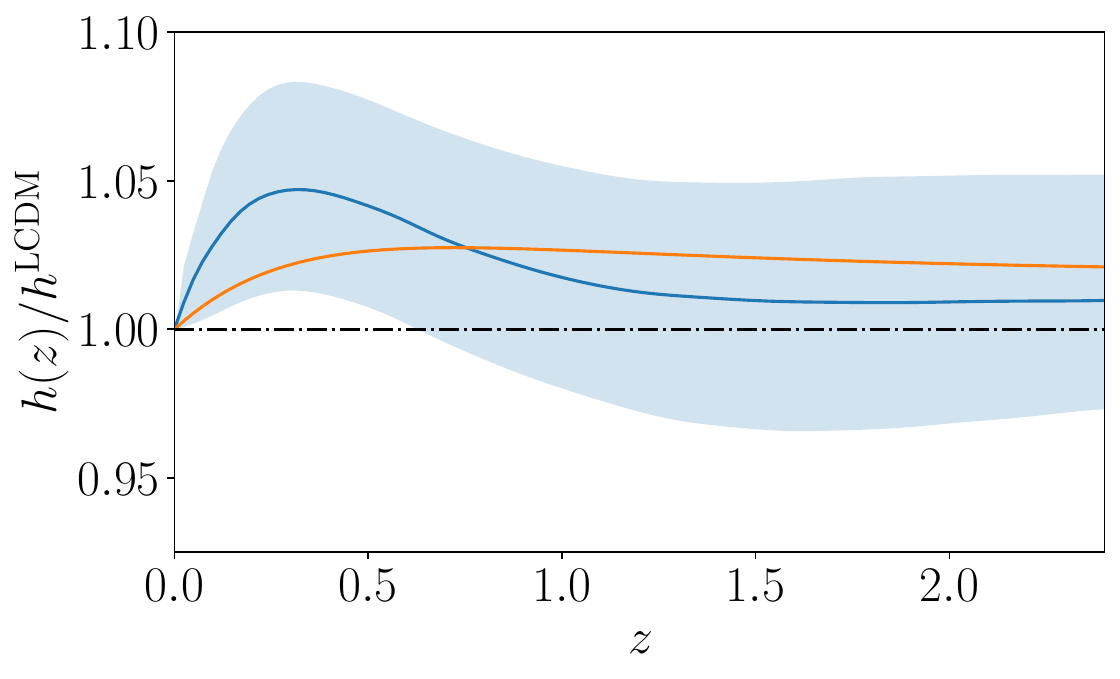}
    \caption{Evolution of $w(z)$ and $h(z)\equiv H(z)/H_0$ for the best-fit $q$CDM$+\Omega_k$ model compared against the DESI reconstruction using CMB+DESI+Union3 data. The shaded regions represent the 95\% confidence regions for the reconstruction (data taken from fig.~1 of~\cite{Calderon:2024uwn}). Note that the $w_0 w_a$ evolution matches the reconstruction very well, so we do not plot it separately.  At $z=0$ we have $w_0 = -0.89$ for $q$CDM$+\Omega_k$ while for the reconstruction it is $w_0=-0.65$.}
    \label{fig:bestfit_comparison}
\end{figure}

We also compare our results against the dark energy model-independent reconstruction of $w(z)$ using Chebyshev polynomials carried out by the DESI collaboration~\cite{Calderon:2024uwn}. To do this, we plot $w(z)$ and the quantity $h(z)\equiv H(z)/H_0$ for the best-fit $q$CDM model vs the DESI reconstruction for the CMB+DESI+Union3 data combination in~\cref{fig:bestfit_comparison}.  We see that even though $w$ does not go below $-1$, the evolution of the Hubble parameter in the best-fit $q$CDM model matches fairly well with the DESI reconstruction. Although the deviation increases at \mbox{$z\approx 0.5$}, the best-fit $q$CDM quantities still lie with the 95\% confidence regions of the DESI reconstruction. It is also clear from the plots that the best-fit $q$CDM model leads to a much lower deviation from $\Lambda$CDM compared to the reconstruction (or for that matter, the $w_0w_a$ parametrisation).

\section{Discussion}\label{sec:discussion}

In this work we have performed a cosmological analysis of single field quintessence for a string theory motivated  potential, namely the exponential $V=V_0 e^{-\lambda\phi}$, with open, closed or flat 3D spaces. 
Using the most recent cosmological data from  
\textit{Planck} CMB, DESI BAO, as well as recent supernovae catalogues, we obtain the following marginalised means and limits on the potential parameter at the exponent:  
\[\lambda = 0.48^{+0.28}_{-0.21},\quad 0.68^{+0.31}_{-0.20},\quad 0.77^{+0.18}_{-0.15}\] 
at 68\% C.L.,  by combining CMB+DESI with PantheonPlus, Union3 and DES-Y5 supernovae datasets respectively. These results indicate an increasing  preference for a non-zero value of $\lambda$, where $\lambda =0$ corresponds to a pure cosmological constant.  Regarding the curvature of the 3D space slices, we find that the results point towards an open universe, $k<0$, with the following marginalised means and limits:
\[\Omega_k= 0.0025\pm 0.0015, \quad 0.0028^{+0.0016}_{-0.0019}, \quad 0.0027\pm 0.0016    \]
at 68\% C.L., with datasets as above; the preference for non-zero curvature is, however, so far not statistically significant.

Whilst exponential quintessence is marginally favoured over $\Lambda$CDM by the cosmological data, our results indicate that the typical values for the parameter $\lambda$ arising in string theory examples, conjectured to be always greater than $\sqrt{2}$ \cite{Bedroya:2019snp, Rudelius:2021oaz} (see \cite{Marconnet:2022fmx,Andriot:2023wvg,Andriot:2024jsh} for constructions with e.g. $\lambda=\sqrt{8/3}, \sqrt{3}$), are excluded. At the same time, given a best-fit value for $\lambda$, the  dynamical system analysis that we presented (see sections \ref{sec:dynsys} and Appendix \ref{app1}) reveals the future evolution of the universe, in the case that dark energy is described by exponential quintessence. The observationally preferred values for $\lambda$, having $\lambda<\sqrt{2}$, indicate that -- rather than having a transient phase of acceleration -- the universe will evolve towards the eternally accelerating fixed point ($P_{\phi}$ or  $\bar Q_{\phi}$), thus giving rise to the presence of a cosmological horizon. Alternatively,   this may well indicate that the simplest single field exponential potentials are not  the correct ones to describe the observed dark energy.

We have also compared our results with the CPL parametrisation used in (past and) recent  cosmological studies of dynamical dark energy~\cite{DESI:2024mwx}, as well as dark energy model independent reconstructions~\cite{Calderon:2024uwn} of $w(z)$ using DESI and supernovae data. The data appear to prefer a more pronounced evolution of $w(z)$ than what can be obtained in the $q$CDM model, requiring $w(z)<-1$ in the past $(z > 1)$, and rapidly growing to $w_0>-1$ closer to the present. In contrast, for the exponential quintessence model, $w(z)$ remains very close to (but always greater than)  $-1$ in the past, and evolves more slowly towards $w_0>-1$ near the present. This results in a mild to moderate preference for the CPL parametrisation over our model, even though the evolution of the background quantities such as $H(z)$ in our curved exponential quintessence model is not significantly different, and still lies within the 95\% confidence regions of the DESI reconstructions~(see \cref{fig:bestfit_comparison}). 

The hint from \cite{Calderon:2024uwn} for  a phantom like equation of state for dark energy in the past is interesting from a fundamental point of view.
In principle, scalar potentials arising from string theory, such as exponentials and hilltops, do all obey the  energy conditions, and therefore do not give rise to phantom behaviour in a consistent way (although  see \cite{Shlivko:2024llw}). However, it would be interesting  to further investigate  whether there can be string  inspired constructions   giving rise to an ``effective" phantom behaviour.
More generally, we emphasize the importance of developing  complete fundamental scenarios that encompass the standard model as well as the dark sectors, in order to address the crucial challenges that any quintessence model  presents, including time-evolution of fundamental constants, unobserved fifth forces, and the ultraviolet stability of the quintessence potential.

As we mentioned before, the present results disfavour the simplest potentials motivated by string theory constructions. On the other hand, it is fair to say that a single exponential is a rather simple case and more complex runaway exponentials could perform better, although see \cite{ValeixoBento:2020ujr} for a no-go on string motivated potentials in supergravity. 
Moreover,  string theory motivated  hilltop potentials \cite{Olguin-Trejo:2018zun} may also perform better (see e.g.~\cite{Shlivko:2024llw}).  Certainly, much work is left 
for  future studies towards systematically comparing string theory models
of cosmology with current and forthcoming cosmological data.

Given that DESI will collect data over a period of five years and that the results discussed here are only based on the first year of observations, there is hope to further improve the comparison between $q$CDM models and the phenomenological CPL parametrisation.  It would also be interesting to compare these models with other well-motivated scenarios of quintessence. At this stage it is too early to say  what was (and will be!) the evolution of the dark energy equation of state parameter in our universe. Hopefully, future data from DESI as well as other dark energy focused experiments will further tighten the constraints on dark energy evolution and allow us to draw   definitive conclusions on the nature of dark energy. 

\section*{Acknowledgements}

We thank Saba Rahimy for discussions. 
The work of GB, AM, GT and IZ is partially funded by  STFC grant ST/X000648/1 and the work of SP is partially funded by STFC grant ST/X000699/1. 
For the initial part of the project, S.B.~was supported by “Progetto di Eccellenza” of the Department of Physics and Astronomy of the University of Padua, Italy. We also acknowledge the support of the Supercomputing Wales project, which is part-funded by the
European Regional Development Fund (ERDF) via Welsh Government. 
For the purpose of
open access, the authors have applied a Creative Commons Attribution licence to any Author
Accepted Manuscript version arising.

\newpage

\begin{appendix}
\section{Open quintessence}\label{app1}
In this appendix we provide a short summary of the open quintessence case, which is described in detail in \cite{Andriot:2024jsh}. The cosmological equations of motion,  \eqref{eq:eoms}, can be written as a dynamical system in terms of the following variables \cite{Andriot:2024jsh}: 
\be\label{eq:variables}
   x= \frac{\phi'}{\sqrt{6}} \,, \qquad 
   y = \frac{\sqrt{V}}{\sqrt{3} H}\,,    \qquad   
   u=\frac{\sqrt{\rho_r}}{\sqrt{3}H}\,,\qquad
   z = \frac{\sqrt{-k}}{a H}\,,  \qquad 
   \lambda = -\frac{V_\phi}{V} \,,
   \ee
together with the constraint 
\be
\Omega_m=1-x^2-y^2-u^2-z^2\,,
\ee
 where prime $'$  denotes derivative with respect to the number of efolds $\d N= H \d t$, as 
\begin{subequations}\label{eq:system1}
 \begin{empheq}[box=\widefbox]{align}
    x'&= \sqrt{\frac{3}{2}}\, y^2\,\lambda + x \left( 3\,(x^2-1)  +  z^2 +\frac32\Omega_m +2u^2\right) \,,\\
  y' &= y \left(- \sqrt{\frac{3}{2}}\, x \,\lambda + 3\,x^2 + z^2 +\frac32 \Omega_m +2u^2\right) \,,\\
  z' &= \,z\left( z^2-1 +3 \, x^2 + \frac32 \Omega_m +2u^2\right)\,,\\
  u' &= \,u\left( z^2-2 +3 \, x^2 + \frac32 \Omega_m +2u^2\right)\,,\\
  \lambda' &= -\sqrt{6}\,x\, \left(\frac{\del_{\phi}^2 V}{V} -\frac{(\del_{\phi}V)^2}{V^2}\right)\,,
\end{empheq}
\end{subequations}
The fixed points for this system are given in  table \ref{tab:3} (see table 2 from \cite{Andriot:2024jsh}).
The properties of the fixed points are as follows \cite{Andriot:2024jsh}:
\begin{itemize}
    \item $P_{kin}$ --  {\sl Kinetic domination}: The energy density is  dominated by the kinetic energy of the scalar field, with  $w_{\rm eff} =1$. These points are the only fully unstable points in the past. 

    \item $P_k$ -- {\sl Curvature domination}: The energy density is dominated by the curvature  with $\Omega_k=1$ and $w_{\rm eff} =-1/3$. This point is a saddle. 

    \item $P_{k\phi}$ -- {\sl Curvature scaling}: At this point, the universe evolves under the influence of both the curvature and the scalar field. However, the expansion mimics pure curvature domination with $w_{\rm eff} =-1/3$. This point exists for $\lambda>\sqrt{2}$ and it is fully stable. 

    \item $P_{\phi}$ -- {\sl Scalar domination}: The energy density is fully  dominated by the scalar field with $\Omega_\phi=1$. This is a standard point that arises in quintessence models  \cite{Bahamonde:2017ize}. It exists for $\lambda<\sqrt{6}$ and  is the only point that allows  acceleration for $\lambda<\sqrt{2}$ with $w_{\rm eff}<-1/3$. For this value of $\lambda$, it is a stable point. It becomes a saddle for $\lambda >\sqrt{2}$.

    \item $P_{m\phi}$ -- {\sl Matter scaling}: The universe evolves under the influence of both matter and  scalar field. Similar to the scaling curvature point, the evolution mimics a matter dominated epoch with $w_{\rm eff}=0$. This point is a saddle. 

    \item $P_m$-- {\sl Matter domination}: The energy density is dominated by matter with $\Omega_m=1$ and  $w_{\rm eff}=0$ and it is a saddle. 

    \item $P_r$ -- {\sl Radiation domination}: The energy density is dominated by radiation with $\Omega_r=1$ and  $w_{\rm eff}=\frac{1}{3}$. This point is also a saddle.

    \item $P_{r\phi}$ -- {\sl Radiation scaling}: The   universe evolves under influence of  radiation and the scalar field. As the other scaling points, it mimics a pure radiation domination universe with $w_{\rm eff}=\frac{1}{3}$. 

\end{itemize}

\begin{table}[H]
\begin{center}
\centering
\begin{tabular}{| l | c | c | c | }
\hline
%\cellcolor[gray]{0.9} & \cellcolor[gray]{0.9} & \cellcolor[gray]{0.9} & \cellcolor[gray]{0.9}\\[-8pt]
\cellcolor[gray]{0.9} \hskip 1.4cm $(x,y,z,u)$ & \cellcolor[gray]{0.9} $\Omega_m$ &  \cellcolor[gray]{0.9} Existence & \cellcolor[gray]{0.9} $w_{\rm eff}$ \\[5pt]
\hline
&&&\\[-8pt]
$P_{\rm kin}= $ $(\pm 1,0,0,0) $ & $0$  & $\forall\, \lambda$ & $1$ \\[3pt]
\hline
&&&\\[-8pt]
$P_{k}= $ $(0,0,\pm1,0) $ & $0$  & $ \forall\, \lambda$ & $ -\frac{1}{3}$ \\[3pt]
\hline
&&&\\[-8pt]
$P_{k\phi}= $ $\lp \frac{1}{\lambda}\sqrt{\frac{2}{3}},\pm\frac{2}{\lambda\sqrt{3}} ,\pm \sqrt{1-\frac{2}{\lambda^2}},0\rp $ & $0$ & $ \lambda>\sqrt{2}$ & $ -\frac{1}{3}$ \\[8pt]
%&&&\\[-8pt]
 & & (For $\lambda=\sqrt{2}$, $P_{k\phi}=P_\phi$) & \\[8pt]
\hline
&&&\\[-8pt]
$P_{\phi}= $ $\lp \frac{\lambda}{\sqrt{6}}, \pm \frac{\sqrt{6-\lambda^2}}{\sqrt{6}}, 0 , 0\rp $ & $0$ & $  \lambda < \sqrt{6}$ & $ \frac{\lambda^2}{3}-1$ \\[8pt]
%&&&\\[-8pt]
 & & (For $\lambda=\sqrt{6}$, $P_{\phi}=P_{kin}$) & \\[8pt]
\hline
&&&\\[-8pt]
$P_{m\phi}= $ $\lp \frac{1}{\lambda} \sqrt{\frac{3}{2}},\pm\frac{1}{\lambda} \sqrt{\frac{3}{2}},0,0\rp $ & $1-\frac{3}{\lambda^2}$ & $  \lambda > \sqrt{3}$  & $0$\\[8pt]
%&&&\\[-8pt]
 & & (For $\lambda=\sqrt{3}$, $P_{m\phi}=P_\phi$) & \\[8pt]
\hline
&&&\\[-8pt]
$P_{m}= $ $(0,0,0,0) $ & $1$ & $ \forall\, \lambda$ & $0$ \\[3pt]
\hline
&&&\\[-8pt]
$P_{r}= $ $(0,0,0,\pm 1)$ & $0$ & $ \forall\, \lambda$ & $\frac{1}{3}$ \\[3pt]
\hline
&&&\\[-8pt]
$P_{r\phi}= $ $\left( \frac{1}{\lambda} \sqrt{\frac{8}{3}},\pm \frac{2}{\lambda\, \sqrt{3}},0, \pm \sqrt{1-\frac{4}{\lambda^2}} \right)$ & $0$ & $\lambda > 2$ & $\frac{1}{3}$ \\[8pt]
%&&&\\[-8pt]
 & & (For $\lambda=2$, $P_{r\phi}=P_\phi$) & \\[8pt]
\hline
\end{tabular}
\end{center}
\caption{Fixed points for the system \eqref{eq:system1} (see \cite{Andriot:2024jsh}).}
\label{tab:3}
\end{table}

The stability of the fixed points in the open quintessence can be found in  table 3 of \cite{Andriot:2024jsh}, which we reproduce in our table \ref{tab:4}.

\begin{table}[H]
\begin{center}
\centering
\begin{tabular}{| l | c | c | c |}
\hline
%\cellcolor[gray]{0.9} & \cellcolor[gray]{0.9} & \cellcolor[gray]{0.9} &  \cellcolor[gray]{0.9} \\[-8pt]
\cellcolor[gray]{0.9} Point & \cellcolor[gray]{0.9} Eigenvalues &  \cellcolor[gray]{0.9} Stability &  \cellcolor[gray]{0.9} Existence \\[5pt]
\hline
%&&&\\[-10pt]
&& $P_{kin}^+:$ Fully unstable \phantom{for $\lambda \leq \sqrt{6}$} & \\
$P_{kin}^\pm $ & $ \lp 3,2,3 \mp \lambda \sqrt{\frac32}, 1 \rp$  & $P_{kin}^-:$ Fully unstable for $\lambda \leq \sqrt{6}$ & - \\
&& $P_{kin}^-:$ Saddle for $\lambda >\sqrt{6}$ \phantom{Fully i} &\\[6pt]
\hline
&&&\\[-8pt]
$P_{k} $ & $(-2,-1,1,-1)$  & Saddle & - \\[4pt]
\hline
&&&\\[-10pt]
$P_{k\phi}$ & $\lp -1, -1 -\frac{\sqrt{8-3\lambda^2}}{\lambda},-1 +\frac{\sqrt{8-3\lambda^2}}{\lambda},-1\rp $  & Stable & $\lambda > \sqrt{2}$ \\[6pt]
\hline
%&&&\\[-7pt]
 &  & Stable for $\lambda < \sqrt{2}$ & \\[-8pt]
$P_\phi$ & $ \lp \frac{\lambda^2}{2}-3, \lambda^2-3,\frac{\lambda^2}{2}-1 , \frac{\lambda^2}{2}-2\rp$ && $\lambda < \sqrt{6}$ \\[-10pt]
&& Saddle for $\lambda > \sqrt{2}$ & \\[6pt]
\hline
&&&\\[-10pt]
$P_{m\phi} $ & $\lp \frac12,-\frac{3(\lambda+\sqrt{24-7\lambda^2})}{4\lambda},-\frac{3(\lambda -\sqrt{24-7\lambda^2})}{4\lambda}, -\frac12 \rp $  & Saddle & $\lambda > \sqrt{3} $ \\[6pt]
\hline
&&&\\[-8pt]
$P_m $ & $ (-\frac32,\frac32,\frac12,-\frac12)$  & Saddle & - \\[5pt]
\hline
&&&\\[-8pt]
$P_{r} $ & $ (-1,2,1,1)$  & Saddle & - \\[5pt]
\hline
&&&\\[-8pt]
$P_{r\phi} $ & $ (1,1, -\frac{\lambda+\sqrt{64-15 \lambda^2}}{2\lambda}, -\frac{\lambda-\sqrt{64-15 \lambda^2}}{2\lambda} )$  & Saddle & $\lambda > 2 $ \\[5pt]
\hline
\end{tabular}
\end{center}
\caption {Stability of the fixed points in table \ref{tab:3} (see reference \cite{Andriot:2024jsh} for details). }
\label{tab:4}
\end{table}

\newpage
\section{Full cosmological constraints}
\label{app:full_constraints}
\begin{figure}[H]
    \centering
    \includegraphics[width=0.95\linewidth]{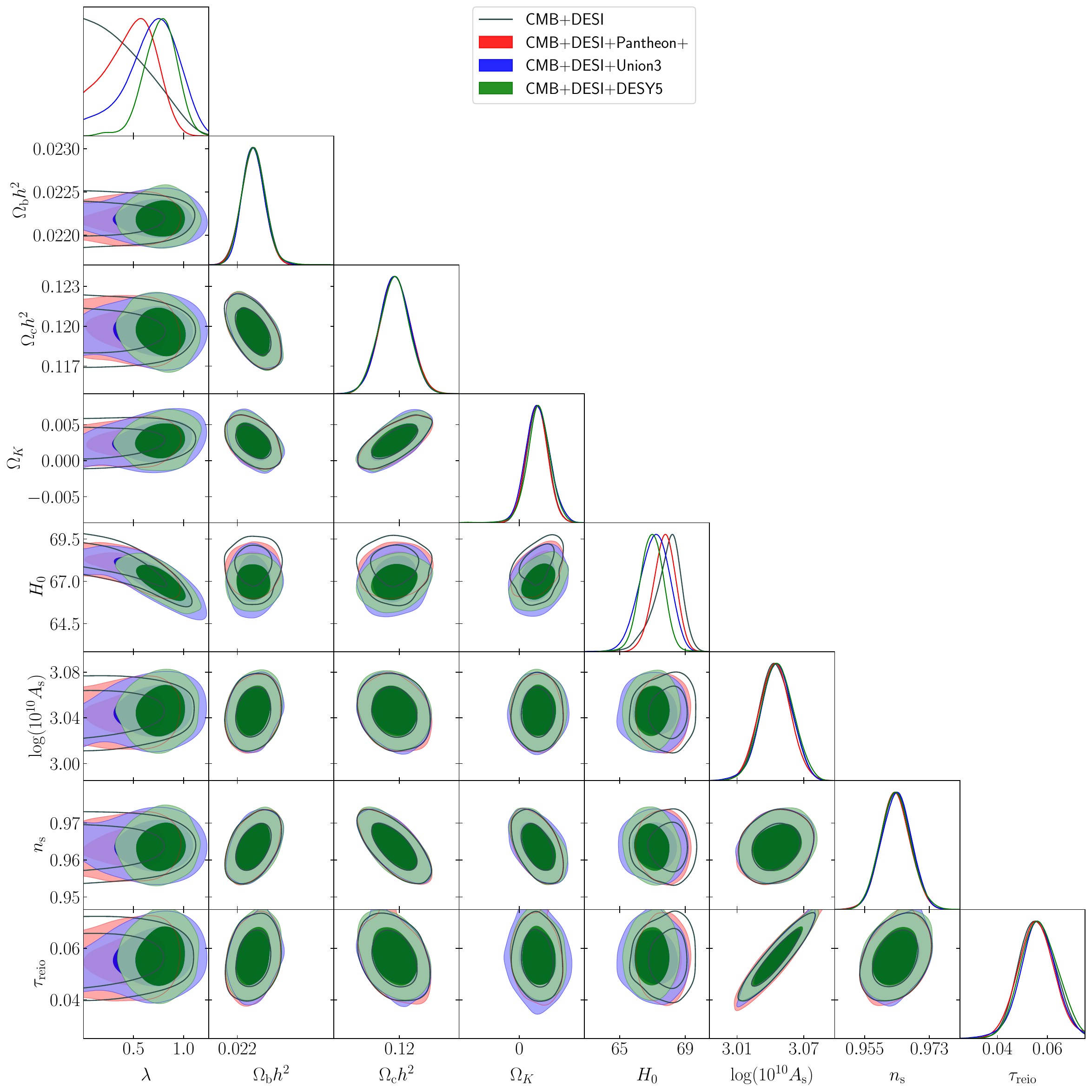}
    \caption{Constraints on the full set of  cosmological parameters for CMB+DESI as well as with the addition of the different supernovae datasets.}
    \label{fig:contours_allparam}
\end{figure}

\begin{figure}[H]
    \centering
    \includegraphics[width=0.95\linewidth]{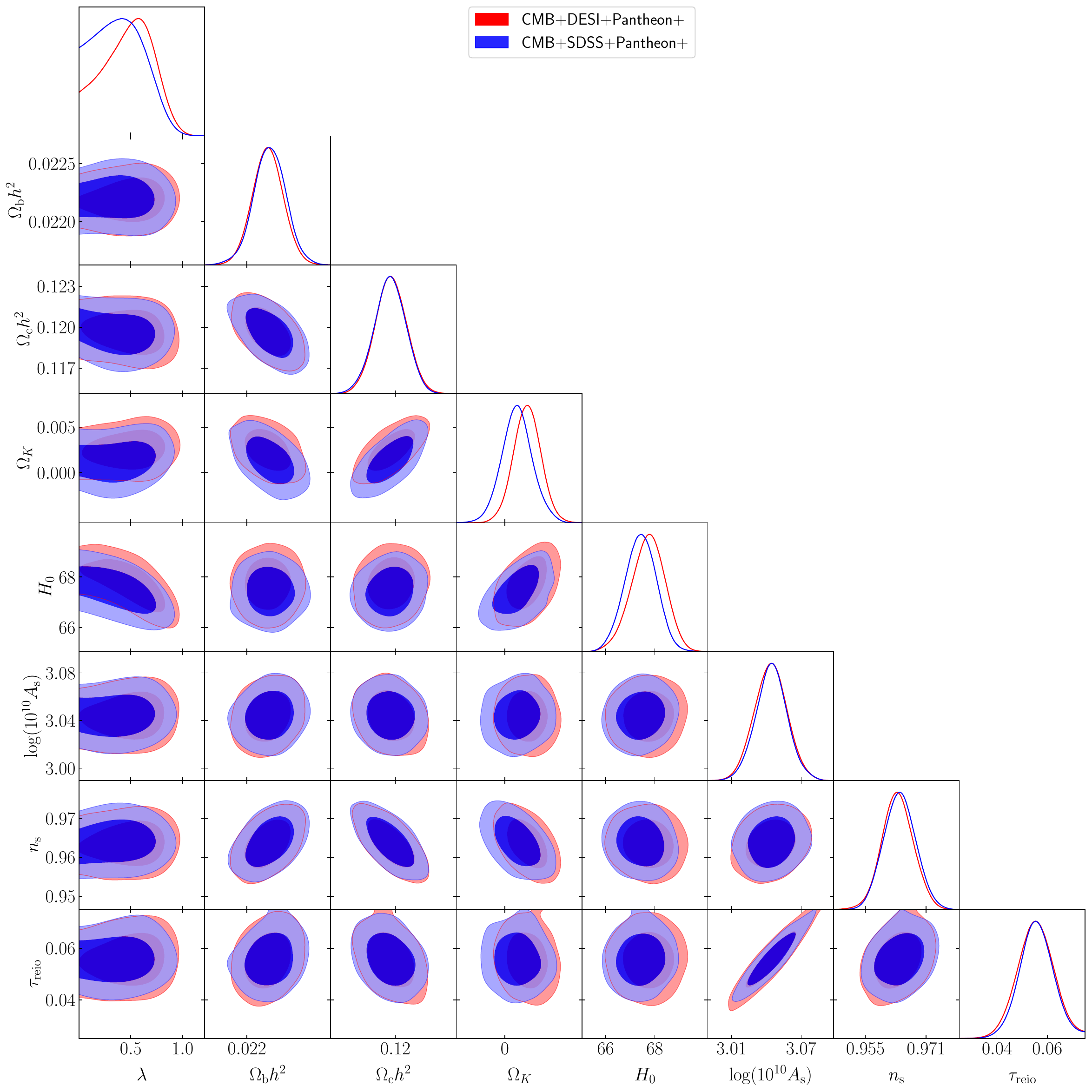}
    \caption{Constraints on cosmological parameters for CMB+DESI+Pantheon+ and CMB+SDSS+Pantheon+.}
    \label{fig:allparam_diffbao}
\end{figure}

\begin{figure}[H]
    \centering
    \includegraphics[width=0.95\linewidth]{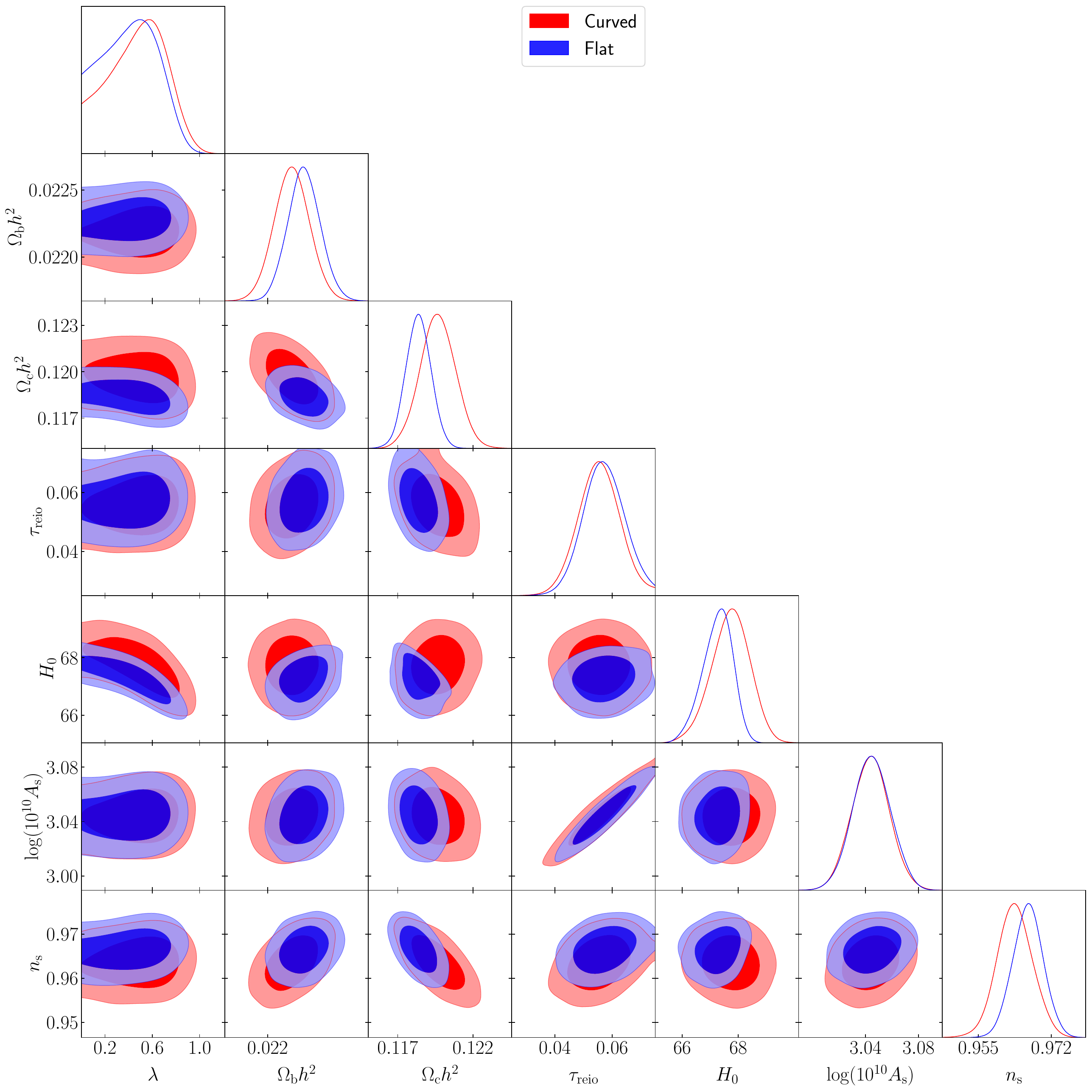}
    \caption{Comparison of the exponential quintessence model analysed in the main text with curved vs flat spatial geometry  for CMB+DESI+Pantheon+. The constraints are slightly relaxed for the model with non-zero $\omk$ but there is no major shift in any cosmological parameter. In particular, the marginalised distribution of $\lambda$ does not show a significant difference. Note that for the curved case, the constraints shown are obtained after marginalising over $\omk$.}
    \label{fig:curved_flat}
\end{figure}

In this appendix we collect the results for the constraints on the full set of cosmological parameters in~\cref{fig:contours_allparam} and~\cref{tab:allparam_main} for the CMB+DESI data as well as with the addition of the various supernovae data described before. In \cref{fig:allparam_diffbao} and \cref{tab:allparam_diffbaofull} we compare the constraints on the parameters using CMB+DESI+Pantheon+ vs CMB+SDSS+Pantheon+. Finally, in~\cref{fig:curved_flat} we show the constraints for the exponential quintessence model under the assumption of flat spatial geometry of the universe. 

\newpage
\begin{table}[H]
    \centering
\begin{tabular} { |c| c| c| c| c|}
\hline
\rowcolor{gray!30} 
 Parameter &  {\bf CMB+DESI} & {\bf +Pantheon+} &  {\bf +Union3+} &  {\bf +DESY5} \\
\hline
{\boldmath$\lambda        $} & $< 0.537                   $ & $0.48^{+0.28}_{-0.21}      $ & $0.68^{+0.31}_{-0.20}      $ & $0.77^{+0.18}_{-0.15}$\\

{\boldmath$\Omega_k       $} & $0.0026\pm 0.0015          $ & $0.0025\pm 0.0015          $ & $0.0028^{+0.0016}_{-0.0019}$ & $0.0027\pm 0.0016          $\\

{\boldmath$\Omega_\mathrm{c} h^2$} & $0.1196\pm 0.0012          $ & $0.1197\pm 0.0012          $ & $0.1195\pm 0.0012          $ & $0.1195\pm 0.0012          $\\

{\boldmath$\log(10^{10} A_\mathrm{s})$} & $3.045\pm 0.014            $ & $3.044\pm 0.014            $ & $3.049\pm 0.013            $ & $3.047\pm 0.014            $\\

{\boldmath$n_\mathrm{s}   $} & $0.9636\pm 0.0041          $ & $0.9635\pm 0.0041          $ & $0.9640\pm 0.0041          $ & $0.9637\pm 0.0047          $\\

{\boldmath$H_0            $} & $67.89^{+0.96}_{-0.61}     $ & $67.73^{+0.72}_{-0.64}     $ & $67.12^{+0.97}_{-0.83}             $ & $66.95\pm 0.72             $\\

{\boldmath$\Omega_\mathrm{b} h^2$} & $0.02219\pm 0.00014        $ & $0.02219\pm 0.00013        $ & $0.02220^{+0.00013}_{-0.00015}$ & $0.02221\pm 0.00013        $\\

{\boldmath$\tau_\mathrm{reio}$} & $0.0559\pm 0.0071          $ & $0.0554\pm 0.0072          $ & $0.0571\pm 0.0067          $ & $0.0577\pm 0.0069          $\\
\hline
\end{tabular}

    \caption{Parameter means and 68\% limits.}
    \label{tab:allparam_main}
\end{table}

\begin{table}[H]
    \centering
\begin{tabular} { |c |c|c|}

\hline
\rowcolor{gray!30} 
\bf{Parameter} & {\bf CMB+DESI+Pantheon+} & {\bf CMB+SDSS+Pantheon+} \\
\hline
{\boldmath$\lambda        $} & $0.48^{+0.28}_{-0.21}      $ & $0.40^{+0.20}_{-0.29}      $\\

{\boldmath$\Omega_k      $} & $0.0025\pm 0.0015          $ & $0.0014\pm 0.0017         $\\

{\boldmath$\Omega_\mathrm{c} h^2$} & $0.1197\pm 0.0012          $ & $0.1197^{+0.0013}_{-0.0012}$\\

{\boldmath$\log(10^{10} A_\mathrm{s})$} & $3.044\pm 0.014            $ & $3.046\pm 0.014            $\\

{\boldmath$n_\mathrm{s}   $} & $0.9635\pm 0.0041          $ & $0.9639\pm 0.0043          $\\

{\boldmath$H_0            $} & $67.73^{+0.72}_{-0.64}     $ & $67.44\pm 0.64     $\\

{\boldmath$\Omega_\mathrm{b} h^2$} & $0.02219\pm 0.00013        $ & $0.02220^{+0.00013}_{-0.00016}$\\

{\boldmath$\tau_\mathrm{reio}$} & $0.0554\pm 0.0072          $ & $0.0562\pm 0.0065          $\\
\hline
\end{tabular}
\caption{Parameter means and 68\% limits.}
    \label{tab:allparam_diffbaofull}
\end{table}

\newpage

\newpage

\end{appendix}

\newpage

\addcontentsline{toc}{section}{References}
\bibliographystyle{utphys}

\bibliography{refsOQC}

\providecommand{\href}[2]{#2}\begingroup\raggedright\begin{thebibliography}{10}

\bibitem{SupernovaSearchTeam:1998fmf}
{\bfseries Supernova Search Team} Collaboration, A.~G. Riess {\em et~al.}, ``{Observational evidence from supernovae for an accelerating universe and a cosmological constant},'' \href{https://dx.doi.org/10.1086/300499}{{\em Astron. J.} {\bfseries 116} (1998) 1009--1038}, \href{https://arxiv.org/abs/astro-ph/9805201}{{\ttfamily arXiv:astro-ph/9805201}}.

\bibitem{SupernovaCosmologyProject:1998vns}
{\bfseries Supernova Cosmology Project} Collaboration, S.~Perlmutter {\em et~al.}, ``{Measurements of $\Omega$ and $\Lambda$ from 42 High Redshift Supernovae},'' \href{https://dx.doi.org/10.1086/307221}{{\em Astrophys. J.} {\bfseries 517} (1999) 565--586}, \href{https://arxiv.org/abs/astro-ph/9812133}{{\ttfamily arXiv:astro-ph/9812133}}.

\bibitem{Abdalla:2022yfr}
E.~Abdalla {\em et~al.}, ``{Cosmology intertwined: A review of the particle physics, astrophysics, and cosmology associated with the cosmological tensions and anomalies},'' \href{https://dx.doi.org/10.1016/j.jheap.2022.04.002}{{\em JHEAp} {\bfseries 34} (2022) 49--211}, \href{https://arxiv.org/abs/2203.06142}{{\ttfamily arXiv:2203.06142 [astro-ph.CO]}}.

\bibitem{Fischler:2001yj}
W.~Fischler, A.~Kashani-Poor, R.~McNees, and S.~Paban, ``{The Acceleration of the universe, a challenge for string theory},'' \href{https://dx.doi.org/10.1088/1126-6708/2001/07/003}{{\em JHEP} {\bfseries 07} (2001) 003}, \href{https://arxiv.org/abs/hep-th/0104181}{{\ttfamily arXiv:hep-th/0104181}}.

\bibitem{Hellerman:2001yi}
S.~Hellerman, N.~Kaloper, and L.~Susskind, ``{String theory and quintessence},'' \href{https://dx.doi.org/10.1088/1126-6708/2001/06/003}{{\em JHEP} {\bfseries 06} (2001) 003}, \href{https://arxiv.org/abs/hep-th/0104180}{{\ttfamily arXiv:hep-th/0104180}}.

\bibitem{Cicoli:2023opf}
M.~Cicoli, J.~P. Conlon, A.~Maharana, S.~Parameswaran, F.~Quevedo, and I.~Zavala, ``{String cosmology: From the early universe to today},'' \href{https://dx.doi.org/10.1016/j.physrep.2024.01.002}{{\em Phys. Rept.} {\bfseries 1059} (2024) 1--155}, \href{https://arxiv.org/abs/2303.04819}{{\ttfamily arXiv:2303.04819 [hep-th]}}.

\bibitem{Garg:2018reu}
S.~K. Garg and C.~Krishnan, ``{Bounds on Slow Roll and the de Sitter Swampland},'' \href{https://dx.doi.org/10.1007/JHEP11(2019)075}{{\em JHEP} {\bfseries 11} (2019) 075}, \href{https://arxiv.org/abs/1807.05193}{{\ttfamily arXiv:1807.05193 [hep-th]}}.

\bibitem{Ooguri:2018wrx}
H.~Ooguri, E.~Palti, G.~Shiu, and C.~Vafa, ``{Distance and de Sitter Conjectures on the Swampland},'' \href{https://dx.doi.org/10.1016/j.physletb.2018.11.018}{{\em Phys. Lett. B} {\bfseries 788} (2019) 180--184}, \href{https://arxiv.org/abs/1810.05506}{{\ttfamily arXiv:1810.05506 [hep-th]}}.

\bibitem{Ratra:1987rm}
B.~Ratra and P.~J.~E. Peebles, ``{Cosmological Consequences of a Rolling Homogeneous Scalar Field},'' \href{https://dx.doi.org/10.1103/PhysRevD.37.3406}{{\em Phys. Rev. D} {\bfseries 37} (1988) 3406}.

\bibitem{Peebles:1987ek}
P.~J.~E. Peebles and B.~Ratra, ``{Cosmology with a Time Variable Cosmological Constant},'' \href{https://dx.doi.org/10.1086/185100}{{\em Astrophys. J. Lett.} {\bfseries 325} (1988) L17}.

\bibitem{Caldwell:1997ii}
R.~R. Caldwell, R.~Dave, and P.~J. Steinhardt, ``{Cosmological imprint of an energy component with general equation of state},'' \href{https://dx.doi.org/10.1103/PhysRevLett.80.1582}{{\em Phys. Rev. Lett.} {\bfseries 80} (1998) 1582--1585}, \href{https://arxiv.org/abs/astro-ph/9708069}{{\ttfamily arXiv:astro-ph/9708069}}.

\bibitem{Copeland:1997et}
E.~J. Copeland, A.~R. Liddle, and D.~Wands, ``{Exponential potentials and cosmological scaling solutions},'' \href{https://dx.doi.org/10.1103/PhysRevD.57.4686}{{\em Phys. Rev. D} {\bfseries 57} (1998) 4686--4690}, \href{https://arxiv.org/abs/gr-qc/9711068}{{\ttfamily arXiv:gr-qc/9711068}}.

\bibitem{Cline:2001nq}
J.~M. Cline, ``{Quintessence, cosmological horizons, and self-tuning},'' \href{https://dx.doi.org/10.1088/1126-6708/2001/08/035}{{\em JHEP} {\bfseries 08} (2001) 035}, \href{https://arxiv.org/abs/hep-ph/0105251}{{\ttfamily arXiv:hep-ph/0105251}}.

\bibitem{Kolda:2001ex}
C.~F. Kolda and W.~Lahneman, ``{Exponential quintessence and the end of acceleration},'' \href{https://arxiv.org/abs/hep-ph/0105300}{{\ttfamily arXiv:hep-ph/0105300}}.

\bibitem{Agrawal:2018own}
P.~Agrawal, G.~Obied, P.~J. Steinhardt, and C.~Vafa, ``{On the Cosmological Implications of the String Swampland},'' \href{https://dx.doi.org/10.1016/j.physletb.2018.07.040}{{\em Phys. Lett. B} {\bfseries 784} (2018) 271--276}, \href{https://arxiv.org/abs/1806.09718}{{\ttfamily arXiv:1806.09718 [hep-th]}}.

\bibitem{Akrami:2018ylq}
Y.~Akrami, R.~Kallosh, A.~Linde, and V.~Vardanyan, ``{The Landscape, the Swampland and the Era of Precision Cosmology},'' \href{https://dx.doi.org/10.1002/prop.201800075}{{\em Fortsch. Phys.} {\bfseries 67} no.~1-2, (2019) 1800075}, \href{https://arxiv.org/abs/1808.09440}{{\ttfamily arXiv:1808.09440 [hep-th]}}.

\bibitem{Raveri:2018ddi}
M.~Raveri, W.~Hu, and S.~Sethi, ``{Swampland Conjectures and Late-Time Cosmology},'' \href{https://dx.doi.org/10.1103/PhysRevD.99.083518}{{\em Phys. Rev. D} {\bfseries 99} no.~8, (2019) 083518}, \href{https://arxiv.org/abs/1812.10448}{{\ttfamily arXiv:1812.10448 [hep-th]}}.

\bibitem{Schoneberg:2023lun}
N.~Sch\"oneberg, L.~Vacher, J.~D.~F. Dias, M.~M. C.~D. Carvalho, and C.~J. A.~P. Martins, ``{News from the Swampland \textemdash{} constraining string theory with astrophysics and cosmology},'' \href{https://dx.doi.org/10.1088/1475-7516/2023/10/039}{{\em JCAP} {\bfseries 10} (2023) 039}, \href{https://arxiv.org/abs/2307.15060}{{\ttfamily arXiv:2307.15060 [astro-ph.CO]}}.

\bibitem{Andriot:2023wvg}
D.~Andriot, D.~Tsimpis, and T.~Wrase, ``{Accelerated expansion of an open universe, and string theory realizations},'' \href{https://arxiv.org/abs/2309.03938}{{\ttfamily arXiv:2309.03938 [hep-th]}}.

\bibitem{Boya:2002mv}
L.~J. Boya, M.~A. Per, and A.~J. Segui, ``{Graphical and kinematical approach to cosmological horizons},'' \href{https://dx.doi.org/10.1103/PhysRevD.66.064009}{{\em Phys. Rev. D} {\bfseries 66} (2002) 064009}, \href{https://arxiv.org/abs/gr-qc/0203074}{{\ttfamily arXiv:gr-qc/0203074}}.

\bibitem{Freivogel:2005vv}
B.~Freivogel, M.~Kleban, M.~Rodriguez~Martinez, and L.~Susskind, ``{Observational consequences of a landscape},'' \href{https://dx.doi.org/10.1088/1126-6708/2006/03/039}{{\em JHEP} {\bfseries 03} (2006) 039}, \href{https://arxiv.org/abs/hep-th/0505232}{{\ttfamily arXiv:hep-th/0505232}}.

\bibitem{Cespedes:2020xpn}
S.~Cespedes, S.~P. de~Alwis, F.~Muia, and F.~Quevedo, ``{Lorentzian vacuum transitions: Open or closed universes?},'' \href{https://dx.doi.org/10.1103/PhysRevD.104.026013}{{\em Phys. Rev. D} {\bfseries 104} no.~2, (2021) 026013}, \href{https://arxiv.org/abs/2011.13936}{{\ttfamily arXiv:2011.13936 [hep-th]}}.

\bibitem{Cespedes:2023jdk}
S.~Cespedes, S.~de~Alwis, F.~Muia, and F.~Quevedo, ``{Quantum Transitions, Detailed Balance, Black Holes and Nothingness},'' \href{https://arxiv.org/abs/2307.13614}{{\ttfamily arXiv:2307.13614 [hep-th]}}.

\bibitem{Bousso:2022gth}
R.~Bousso and E.~Wildenhain, ``{Islands in closed and open universes},'' \href{https://dx.doi.org/10.1103/PhysRevD.105.086012}{{\em Phys. Rev. D} {\bfseries 105} no.~8, (2022) 086012}, \href{https://arxiv.org/abs/2202.05278}{{\ttfamily arXiv:2202.05278 [hep-th]}}.

\bibitem{Ben-Dayan:2022nmb}
I.~Ben-Dayan, M.~Hadad, and E.~Wildenhain, ``{Islands in the fluid: islands are common in cosmology},'' \href{https://dx.doi.org/10.1007/JHEP03(2023)077}{{\em JHEP} {\bfseries 03} (2023) 077}, \href{https://arxiv.org/abs/2211.16600}{{\ttfamily arXiv:2211.16600 [hep-th]}}.

\bibitem{Almheiri:2020cfm}
A.~Almheiri, T.~Hartman, J.~Maldacena, E.~Shaghoulian, and A.~Tajdini, ``{The entropy of Hawking radiation},'' \href{https://dx.doi.org/10.1103/RevModPhys.93.035002}{{\em Rev. Mod. Phys.} {\bfseries 93} no.~3, (2021) 035002}, \href{https://arxiv.org/abs/2006.06872}{{\ttfamily arXiv:2006.06872 [hep-th]}}.

\bibitem{Engelhardt:2014gca}
N.~Engelhardt and A.~C. Wall, ``{Quantum Extremal Surfaces: Holographic Entanglement Entropy beyond the Classical Regime},'' \href{https://dx.doi.org/10.1007/JHEP01(2015)073}{{\em JHEP} {\bfseries 01} (2015) 073}, \href{https://arxiv.org/abs/1408.3203}{{\ttfamily arXiv:1408.3203 [hep-th]}}.

\bibitem{Hartman:2020khs}
T.~Hartman, Y.~Jiang, and E.~Shaghoulian, ``{Islands in cosmology},'' \href{https://dx.doi.org/10.1007/JHEP11(2020)111}{{\em JHEP} {\bfseries 11} (2020) 111}, \href{https://arxiv.org/abs/2008.01022}{{\ttfamily arXiv:2008.01022 [hep-th]}}.

\bibitem{Andriot:2024jsh}
D.~Andriot, S.~Parameswaran, D.~Tsimpis, T.~Wrase, and I.~Zavala, ``{Exponential Quintessence: curved, steep and stringy?},'' \href{https://arxiv.org/abs/2405.09323}{{\ttfamily arXiv:2405.09323 [hep-th]}}.

\bibitem{Bahamonde:2017ize}
S.~Bahamonde, C.~G. B\"ohmer, S.~Carloni, E.~J. Copeland, W.~Fang, and N.~Tamanini, ``{Dynamical systems applied to cosmology: dark energy and modified gravity},'' \href{https://dx.doi.org/10.1016/j.physrep.2018.09.001}{{\em Phys. Rept.} {\bfseries 775-777} (2018) 1--122}, \href{https://arxiv.org/abs/1712.03107}{{\ttfamily arXiv:1712.03107 [gr-qc]}}.

\bibitem{vandenHoogen:1999qq}
R.~J. van~den Hoogen, A.~A. Coley, and D.~Wands, ``{Scaling solutions in Robertson-Walker space-times},'' \href{https://dx.doi.org/10.1088/0264-9381/16/6/317}{{\em Class. Quant. Grav.} {\bfseries 16} (1999) 1843--1851}, \href{https://arxiv.org/abs/gr-qc/9901014}{{\ttfamily arXiv:gr-qc/9901014}}.

\bibitem{Gosenca:2015qha}
M.~Gosenca and P.~Coles, ``{Dynamical Analysis of Scalar Field Cosmologies with Spatial Curvature},'' \href{https://dx.doi.org/10.21105/astro.1502.04020}{{\em Open J. Astrophys.} {\bfseries 1} no.~1, (2016) 1}, \href{https://arxiv.org/abs/1502.04020}{{\ttfamily arXiv:1502.04020 [gr-qc]}}.

\bibitem{Marconnet:2022fmx}
P.~Marconnet and D.~Tsimpis, ``{Universal accelerating cosmologies from 10d supergravity},'' \href{https://dx.doi.org/10.1007/JHEP01(2023)033}{{\em JHEP} {\bfseries 01} (2023) 033}, \href{https://arxiv.org/abs/2210.10813}{{\ttfamily arXiv:2210.10813 [hep-th]}}.

\bibitem{SavasArapoglu:2017pyh}
A.~Sava\c{s}~Arapo\u{g}lu and A.~Emrah~Y\"ukselci, ``{Dynamical System Analysis of Quintessence Models with Exponential Potential - Revisited},'' \href{https://dx.doi.org/10.1142/S021773231950069X}{{\em Mod. Phys. Lett. A} {\bfseries 34} no.~09, (2019) 1950069}, \href{https://arxiv.org/abs/1711.03824}{{\ttfamily arXiv:1711.03824 [gr-qc]}}.

\bibitem{DES:2024tys}
{\bfseries DES} Collaboration, T.~M.~C. Abbott {\em et~al.}, ``{The Dark Energy Survey: Cosmology Results With \textasciitilde{}1500 New High-redshift Type Ia Supernovae Using The Full 5-year Dataset},'' \href{https://arxiv.org/abs/2401.02929}{{\ttfamily arXiv:2401.02929 [astro-ph.CO]}}.

\bibitem{DESI:2024lzq}
{\bfseries DESI} Collaboration, A.~G. Adame {\em et~al.}, ``{DESI 2024 IV: Baryon Acoustic Oscillations from the Lyman Alpha Forest},'' \href{https://arxiv.org/abs/2404.03001}{{\ttfamily arXiv:2404.03001 [astro-ph.CO]}}.

\bibitem{Euclid:2019clj}
{\bfseries Euclid} Collaboration, A.~Blanchard {\em et~al.}, ``{Euclid preparation. VII. Forecast validation for Euclid cosmological probes},'' \href{https://dx.doi.org/10.1051/0004-6361/202038071}{{\em Astron. Astrophys.} {\bfseries 642} (2020) A191}, \href{https://arxiv.org/abs/1910.09273}{{\ttfamily arXiv:1910.09273 [astro-ph.CO]}}.

\bibitem{2009arXiv0912.0201L}
{LSST Science Collaboration}, P.~A. {Abell}, {\em et~al.}, ``{LSST Science Book, Version 2.0}'' \href{https://dx.doi.org/10.48550/arXiv.0912.0201}{{\em arXiv e-prints} (Dec., 2009) arXiv:0912.0201}, \href{https://arxiv.org/abs/0912.0201}{{\ttfamily arXiv:0912.0201 [astro-ph.IM]}}.

\bibitem{DESI:2024uvr}
{\bfseries DESI} Collaboration, A.~G. Adame {\em et~al.}, ``{DESI 2024 III: Baryon Acoustic Oscillations from Galaxies and Quasars},'' \href{https://arxiv.org/abs/2404.03000}{{\ttfamily arXiv:2404.03000 [astro-ph.CO]}}.

\bibitem{DESI:2024mwx}
{\bfseries DESI} Collaboration, A.~G. Adame {\em et~al.}, ``{DESI 2024 VI: Cosmological Constraints from the Measurements of Baryon Acoustic Oscillations},'' \href{https://arxiv.org/abs/2404.03002}{{\ttfamily arXiv:2404.03002 [astro-ph.CO]}}.

\bibitem{Calderon:2024uwn}
R.~Calderon {\em et~al.}, ``{DESI 2024: Reconstructing Dark Energy using Crossing Statistics with DESI DR1 BAO data},'' \href{https://arxiv.org/abs/2405.04216}{{\ttfamily arXiv:2405.04216 [astro-ph.CO]}}.

\bibitem{DESI:2024kob}
{\bfseries DESI} Collaboration, K.~Lodha {\em et~al.}, ``{DESI 2024: Constraints on Physics-Focused Aspects of Dark Energy using DESI DR1 BAO Data},'' \href{https://arxiv.org/abs/2405.13588}{{\ttfamily arXiv:2405.13588 [astro-ph.CO]}}.

\bibitem{Lodha:2024upq}
K.~Lodha {\em et~al.}, ``{DESI 2024: Constraints on Physics-Focused Aspects of Dark Energy using DESI DR1 BAO Data},'' \href{https://arxiv.org/abs/2405.13588}{{\ttfamily arXiv:2405.13588 [astro-ph.CO]}}.

\bibitem{Colgain:2024xqj}
E.~O. Colg\'ain, M.~G. Dainotti, S.~Capozziello, S.~Pourojaghi, M.~M. Sheikh-Jabbari, and D.~Stojkovic, ``{Does DESI 2024 Confirm $\Lambda$CDM?},'' \href{https://arxiv.org/abs/2404.08633}{{\ttfamily arXiv:2404.08633 [astro-ph.CO]}}.

\bibitem{Carloni:2024zpl}
Y.~Carloni, O.~Luongo, and M.~Muccino, ``{Does dark energy really revive using DESI 2024 data?},'' \href{https://arxiv.org/abs/2404.12068}{{\ttfamily arXiv:2404.12068 [astro-ph.CO]}}.

\bibitem{Park:2024jns}
C.-G. Park, J.~de~Cruz~Perez, and B.~Ratra, ``{Using non-DESI data to confirm and strengthen the DESI 2024 spatially-flat $w_0w_a$CDM cosmological parameterization result},'' \href{https://arxiv.org/abs/2405.00502}{{\ttfamily arXiv:2405.00502 [astro-ph.CO]}}.

\bibitem{Wang:2024rjd}
D.~Wang, ``{The Self-Consistency of DESI Analysis and Comment on ''Does DESI 2024 Confirm $\Lambda$CDM?''},'' \href{https://arxiv.org/abs/2404.13833}{{\ttfamily arXiv:2404.13833 [astro-ph.CO]}}.

\bibitem{Cortes:2024lgw}
M.~Cort\^es and A.~R. Liddle, ``{Interpreting DESI's evidence for evolving dark energy},'' \href{https://arxiv.org/abs/2404.08056}{{\ttfamily arXiv:2404.08056 [astro-ph.CO]}}.

\bibitem{Wang:2024pui}
Z.~Wang, S.~Lin, Z.~Ding, and B.~Hu, ``{The role of LRG1 and LRG2's monopole in inferring the DESI 2024 BAO cosmology},'' \href{https://arxiv.org/abs/2405.02168}{{\ttfamily arXiv:2405.02168 [astro-ph.CO]}}.

\bibitem{Dinda:2024kjf}
B.~R. Dinda, ``{A new diagnostic for the null test of dynamical dark energy in light of DESI 2024 and other BAO data},'' \href{https://arxiv.org/abs/2405.06618}{{\ttfamily arXiv:2405.06618 [astro-ph.CO]}}.

\bibitem{Croker:2024jfg}
K.~S. Croker, G.~Tarl\'e, S.~P. Ahlen, B.~G. Cartwright, D.~Farrah, N.~Fernandez, and R.~A. Windhorst, ``{DESI Dark Energy Time Evolution is Recovered by Cosmologically Coupled Black Holes},'' \href{https://arxiv.org/abs/2405.12282}{{\ttfamily arXiv:2405.12282 [astro-ph.CO]}}.

\bibitem{Wang:2024hks}
D.~Wang, ``{Constraining Cosmological Physics with DESI BAO Observations},'' \href{https://arxiv.org/abs/2404.06796}{{\ttfamily arXiv:2404.06796 [astro-ph.CO]}}.

\bibitem{Luongo:2024fww}
O.~Luongo and M.~Muccino, ``{Model independent cosmographic constraints from DESI 2024},'' \href{https://arxiv.org/abs/2404.07070}{{\ttfamily arXiv:2404.07070 [astro-ph.CO]}}.

\bibitem{Mukherjee:2024ryz}
P.~Mukherjee and A.~A. Sen, ``{Model-independent cosmological inference post DESI DR1 BAO measurements},'' \href{https://arxiv.org/abs/2405.19178}{{\ttfamily arXiv:2405.19178 [astro-ph.CO]}}.

\bibitem{Wang:2024dka}
H.~Wang and Y.-S. Piao, ``{Dark energy in light of recent DESI BAO and Hubble tension},'' \href{https://arxiv.org/abs/2404.18579}{{\ttfamily arXiv:2404.18579 [astro-ph.CO]}}.

\bibitem{Tada:2024znt}
Y.~Tada and T.~Terada, ``{Quintessential interpretation of the evolving dark energy in light of DESI},'' \href{https://arxiv.org/abs/2404.05722}{{\ttfamily arXiv:2404.05722 [astro-ph.CO]}}.

\bibitem{Yin:2024hba}
W.~Yin, ``{Cosmic Clues: DESI, Dark Energy, and the Cosmological Constant Problem},'' \href{https://arxiv.org/abs/2404.06444}{{\ttfamily arXiv:2404.06444 [hep-ph]}}.

\bibitem{Berghaus:2024kra}
K.~V. Berghaus, J.~A. Kable, and V.~Miranda, ``{Quantifying Scalar Field Dynamics with DESI 2024 Y1 BAO measurements},'' \href{https://arxiv.org/abs/2404.14341}{{\ttfamily arXiv:2404.14341 [astro-ph.CO]}}.

\bibitem{Shlivko:2024llw}
D.~Shlivko and P.~Steinhardt, ``{Assessing observational constraints on dark energy},'' \href{https://arxiv.org/abs/2405.03933}{{\ttfamily arXiv:2405.03933 [astro-ph.CO]}}.

\bibitem{Chevallier:2000qy}
M.~Chevallier and D.~Polarski, ``{Accelerating universes with scaling dark matter},'' \href{https://dx.doi.org/10.1142/S0218271801000822}{{\em Int. J. Mod. Phys. D} {\bfseries 10} (2001) 213--224}, \href{https://arxiv.org/abs/gr-qc/0009008}{{\ttfamily arXiv:gr-qc/0009008}}.

\bibitem{Linder:2002et}
E.~V. Linder, ``{Exploring the expansion history of the universe},'' \href{https://dx.doi.org/10.1103/PhysRevLett.90.091301}{{\em Phys. Rev. Lett.} {\bfseries 90} (2003) 091301}, \href{https://arxiv.org/abs/astro-ph/0208512}{{\ttfamily arXiv:astro-ph/0208512}}.

\bibitem{Aurich:2003it}
R.~Aurich and F.~Steiner, ``{Quintessence and the curvature of the universe after wmap},'' \href{https://dx.doi.org/10.1142/S0218271804003615}{{\em Int. J. Mod. Phys. D} {\bfseries 13} (2004) 123--136}, \href{https://arxiv.org/abs/astro-ph/0302264}{{\ttfamily arXiv:astro-ph/0302264}}.

\bibitem{Brout:2022vxf}
D.~Brout {\em et~al.}, ``{The Pantheon+ Analysis: Cosmological Constraints},'' \href{https://dx.doi.org/10.3847/1538-4357/ac8e04}{{\em Astrophys. J.} {\bfseries 938} no.~2, (2022) 110}, \href{https://arxiv.org/abs/2202.04077}{{\ttfamily arXiv:2202.04077 [astro-ph.CO]}}.

\bibitem{Rubin:2023ovl}
D.~Rubin {\em et~al.}, ``{Union Through UNITY: Cosmology with 2,000 SNe Using a Unified Bayesian Framework},'' \href{https://arxiv.org/abs/2311.12098}{{\ttfamily arXiv:2311.12098 [astro-ph.CO]}}.

\bibitem{Monteroetal}
G.~Alestas, M.~Delgado, I.~Ruiz, Y.~Akrami, M.~Montero, and S.~Nesseris \href{https://arxiv.org/abs/to appear}{{\ttfamily to appear}}.

\bibitem{Ramadan:2024kmn}
O.~F. Ramadan, J.~Sakstein, and D.~Rubin, ``{DESI Constraints on Exponential Quintessence},'' \href{https://arxiv.org/abs/2405.18747}{{\ttfamily arXiv:2405.18747 [astro-ph.CO]}}.

\bibitem{Lewis:1999bs}
A.~Lewis, A.~Challinor, and A.~Lasenby, ``{Efficient computation of CMB anisotropies in closed FRW models},'' \href{https://dx.doi.org/10.1086/309179}{{\em Astrophys. J.} {\bfseries 538} (2000) 473--476}, \href{https://arxiv.org/abs/astro-ph/9911177}{{\ttfamily arXiv:astro-ph/9911177 [astro-ph]}}.
\url{https://arxiv.org/abs/astro-ph/9911177}.
%%CITATION = ASTRO-PH/9911177;%%.

\bibitem{Howlett:2012mh}
C.~Howlett, A.~Lewis, A.~Hall, and A.~Challinor, ``{CMB power spectrum parameter degeneracies in the era of precision cosmology},'' \href{https://dx.doi.org/10.1088/1475-7516/2012/04/027}{{\em JCAP} {\bfseries 1204} (2012) 027}, \href{https://arxiv.org/abs/1201.3654}{{\ttfamily arXiv:1201.3654 [astro-ph.CO]}}.
\url{https://arxiv.org/abs/1201.3654}.
%%CITATION = ARXIV:1201.3654;%%.

\bibitem{Planck:2019nip}
{\bfseries Planck} Collaboration, N.~Aghanim {\em et~al.}, ``{Planck 2018 results. V. CMB power spectra and likelihoods},'' \href{https://dx.doi.org/10.1051/0004-6361/201936386}{{\em Astron. Astrophys.} {\bfseries 641} (2020) A5}, \href{https://arxiv.org/abs/1907.12875}{{\ttfamily arXiv:1907.12875 [astro-ph.CO]}}.

\bibitem{Aghanim:2019ame}
{\bfseries Planck} Collaboration, N.~Aghanim {\em et~al.}, ``{Planck 2018 results. V. CMB power spectra and likelihoods},'' \href{https://dx.doi.org/10.1051/0004-6361/201936386}{{\em Astron. Astrophys.} {\bfseries 641} (2020) A5}, \href{https://arxiv.org/abs/1907.12875}{{\ttfamily arXiv:1907.12875 [astro-ph.CO]}}.

\bibitem{Rosenberg:2022sdy}
E.~Rosenberg, S.~Gratton, and G.~Efstathiou, ``{CMB power spectra and cosmological parameters from Planck PR4 with CamSpec},'' \href{https://dx.doi.org/10.1093/mnras/stac2744}{{\em Mon. Not. Roy. Astron. Soc.} {\bfseries 517} no.~3, (2022) 4620--4636}, \href{https://arxiv.org/abs/2205.10869}{{\ttfamily arXiv:2205.10869 [astro-ph.CO]}}.

\bibitem{Aghanim:2018oex}
{\bfseries Planck} Collaboration, N.~Aghanim {\em et~al.}, ``{Planck 2018 results. VIII. Gravitational lensing},'' \href{https://dx.doi.org/10.1051/0004-6361/201833886}{{\em Astron. Astrophys.} {\bfseries 641} (2020) A8}, \href{https://arxiv.org/abs/1807.06210}{{\ttfamily arXiv:1807.06210 [astro-ph.CO]}}.

\bibitem{Alam:2020sor}
{\bfseries eBOSS} Collaboration, S.~Alam {\em et~al.}, ``{Completed SDSS-IV extended Baryon Oscillation Spectroscopic Survey: Cosmological implications from two decades of spectroscopic surveys at the Apache Point Observatory},'' \href{https://dx.doi.org/10.1103/PhysRevD.103.083533}{{\em Phys. Rev. D} {\bfseries 103} no.~8, (2021) 083533}, \href{https://arxiv.org/abs/2007.08991}{{\ttfamily arXiv:2007.08991 [astro-ph.CO]}}.

\bibitem{Lewis:2002ah}
A.~Lewis and S.~Bridle, ``{Cosmological parameters from CMB and other data: A Monte Carlo approach},'' \href{https://dx.doi.org/10.1103/PhysRevD.66.103511}{{\em Phys. Rev.} {\bfseries D66} (2002) 103511}, \href{https://arxiv.org/abs/astro-ph/0205436}{{\ttfamily arXiv:astro-ph/0205436 [astro-ph]}}.
\url{https://arxiv.org/abs/astro-ph/0205436}.
%%CITATION = ASTRO-PH/0205436;%%.

\bibitem{Lewis:2013hha}
A.~Lewis, ``{Efficient sampling of fast and slow cosmological parameters},'' \href{https://dx.doi.org/10.1103/PhysRevD.87.103529}{{\em Phys. Rev.} {\bfseries D87} no.~10, (2013) 103529}, \href{https://arxiv.org/abs/1304.4473}{{\ttfamily arXiv:1304.4473 [astro-ph.CO]}}.
\url{https://arxiv.org/abs/1304.4473}.
%%CITATION = ARXIV:1304.4473;%%.

\bibitem{Torrado:2020dgo}
J.~Torrado and A.~Lewis, ``{Cobaya: Code for Bayesian Analysis of hierarchical physical models},'' \href{https://dx.doi.org/10.1088/1475-7516/2021/05/057}{{\em JCAP} {\bfseries 05} (2021) 057}, \href{https://arxiv.org/abs/2005.05290}{{\ttfamily arXiv:2005.05290 [astro-ph.IM]}}.

\bibitem{Lewis:2019xzd}
A.~Lewis, ``{GetDist: a Python package for analysing Monte Carlo samples},'' \href{https://arxiv.org/abs/1910.13970}{{\ttfamily arXiv:1910.13970 [astro-ph.IM]}}.

\bibitem{Bobyqa1}
C.~{Cartis}, L.~{Roberts}, and O.~{Sheridan-Methven}, ``{Escaping local minima with derivative-free methods: a numerical investigation},'' \href{https://dx.doi.org/10.48550/arXiv.1812.11343}{{\em arXiv e-prints} (Dec., 2018) arXiv:1812.11343}, \href{https://arxiv.org/abs/1812.11343}{{\ttfamily arXiv:1812.11343 [math.OC]}}.

\bibitem{Bobyqa2}
C.~{Cartis}, J.~{Fiala}, B.~{Marteau}, and L.~{Roberts}, ``{Improving the Flexibility and Robustness of Model-Based Derivative-Free Optimization Solvers},'' \href{https://dx.doi.org/10.48550/arXiv.1804.00154}{{\em arXiv e-prints} (Mar., 2018) arXiv:1804.00154}, \href{https://arxiv.org/abs/1804.00154}{{\ttfamily arXiv:1804.00154 [math.OC]}}.

\bibitem{Liddle:2004nh}
A.~R. Liddle, ``{How many cosmological parameters?},'' \href{https://dx.doi.org/10.1111/j.1365-2966.2004.08033.x}{{\em Mon. Not. Roy. Astron. Soc.} {\bfseries 351} (2004) L49--L53}, \href{https://arxiv.org/abs/astro-ph/0401198}{{\ttfamily arXiv:astro-ph/0401198}}.

\bibitem{AIC}
H.~{Akaike}, ``{A New Look at the Statistical Model Identification},'' {\em IEEE Transactions on Automatic Control} {\bfseries 19} (Jan., 1974) 716--723.

\bibitem{Bedroya:2019snp}
A.~Bedroya and C.~Vafa, ``{Trans-Planckian Censorship and the Swampland},'' \href{https://dx.doi.org/10.1007/JHEP09(2020)123}{{\em JHEP} {\bfseries 09} (2020) 123}, \href{https://arxiv.org/abs/1909.11063}{{\ttfamily arXiv:1909.11063 [hep-th]}}.

\bibitem{Rudelius:2021oaz}
T.~Rudelius, ``{Dimensional reduction and (Anti) de Sitter bounds},'' \href{https://dx.doi.org/10.1007/JHEP08(2021)041}{{\em JHEP} {\bfseries 08} (2021) 041}, \href{https://arxiv.org/abs/2101.11617}{{\ttfamily arXiv:2101.11617 [hep-th]}}.

\bibitem{ValeixoBento:2020ujr}
B.~Valeixo~Bento, D.~Chakraborty, S.~L. Parameswaran, and I.~Zavala, ``{Dark Energy in String Theory},'' \href{https://dx.doi.org/10.22323/1.376.0123}{{\em PoS} {\bfseries CORFU2019} (2020) 123}, \href{https://arxiv.org/abs/2005.10168}{{\ttfamily arXiv:2005.10168 [hep-th]}}.

\bibitem{Olguin-Trejo:2018zun}
Y.~Olguin-Trejo, S.~L. Parameswaran, G.~Tasinato, and I.~Zavala, ``{Runaway Quintessence, Out of the Swampland},'' \href{https://dx.doi.org/10.1088/1475-7516/2019/01/031}{{\em JCAP} {\bfseries 01} (2019) 031}, \href{https://arxiv.org/abs/1810.08634}{{\ttfamily arXiv:1810.08634 [hep-th]}}.

\end{thebibliography}\endgroup

\end{document}